\documentclass[10pt,journal,compsoc]{IEEEtran}
\ifCLASSOPTIONcompsoc
  \usepackage[nocompress]{cite}
\else
  \usepackage{cite}
\fi
\ifCLASSINFOpdf
\else
\fi
\usepackage{amsmath,amsthm,amsbsy,amssymb,amsfonts,booktabs,subfigure,url,stfloats,graphicx,multirow,wasysym,algpseudocode,flushend,enumitem,url,tcolorbox,hyperref}
\usepackage[ruled,vlined,linesnumbered]{algorithm2e}
\usepackage{soul}
\soulregister\cite7
\soulregister\citep7
\soulregister\citet7
\soulregister\ref7
\soulregister\pageref7
\soulregister\em7

\usepackage{etoolbox}
\makeatletter 
\pretocmd\@bibitem{\csname keycolor#1\endcsname}{}{\fail}
\newcommand\citecolor[3][1]{\@namedef{keycolor#3}{\hspace*{-\labelwidth}\hspace*{-\labelsep}{\color{#2}\rule[-0.3em]{\dimexpr\linewidth+\labelwidth+\labelsep\relax}{#1\baselineskip}}\vspace*{\itemsep}\vspace*{-#1\baselineskip}}}
\makeatother

\usepackage{listings,color,xcolor,colortbl}
\definecolor{dkgreen}{rgb}{0,0.6,0}
\definecolor{gray}{rgb}{0.5,0.5,0.5}
\definecolor{mauve}{rgb}{0.58,0,0.82}
\begin{document}
\title{MalPurifier: Enhancing Android Malware Detection with Adversarial Purification against Evasion Attacks}

 \author{Yuyang~Zhou,~\IEEEmembership{Member,~IEEE},
  ~Guang~Cheng,~\IEEEmembership{Member,~IEEE},
  ~Zongyao~Chen,~\IEEEmembership{Student Member,~IEEE},
  ~and Shui~Yu,~\IEEEmembership{Fellow,~IEEE}
  % <-this % stops a space
  \IEEEcompsocitemizethanks{\IEEEcompsocthanksitem Yuyang Zhou, Guang Cheng, and Zongyao Chen are with the School of Cyber Science and Engineering, Southeast University, Purple Mountain Laboratories, and Jiangsu Province Engineering Research Center of Security for Ubiquitous Network, Nanjing 211189, China. E-mail: \{yyzhou, chengguang\}@seu.edu.cn, zongyao.chen@linux.alibaba.com.

  \IEEEcompsocthanksitem Shui Yu is with the School of Computer Science, University of Technology Sydney, Ultimo, NSW 2007, Australia. E-mail: Shui.Yu@uts.edu.au.

  \IEEEcompsocthanksitem Guang Cheng is the corresponding author.}
}


\markboth{Submitted to IEEE for Peer Review.}%
{Yuyang Zhou, Guang Cheng, Zongyao Chen, and Shui Yu.}



\IEEEtitleabstractindextext{%
\begin{abstract}  
  Machine learning (ML) has gained significant adoption in Android malware detection to address the escalating threats posed by the rapid proliferation of malware attacks. However, recent studies have revealed the inherent vulnerabilities of ML-based detection systems to evasion attacks. While efforts have been made to address this critical issue, many of the existing defensive methods encounter challenges such as lower effectiveness or reduced generalization capabilities. In this paper, we introduce MalPurifier, a novel adversarial purification framework specifically engineered for Android malware detection. Specifically, MalPurifier integrates three key innovations: a diversified adversarial perturbation mechanism for robustness and generalizability, a protective noise injection strategy for benign data integrity, and a Denoising AutoEncoder (DAE) with a dual-objective loss for accurate purification and classification. Extensive experiments on two large-scale datasets demonstrate that MalPurifier significantly outperforms state-of-the-art defenses. It robustly defends against a comprehensive set of 37 perturbation-based evasion attacks, consistently achieving robust accuracies above 90.91\%. As a lightweight, model-agnostic, and plug-and-play module, MalPurifier offers a practical and effective solution to bolster the security of ML-based Android malware detectors.
\end{abstract}

\begin{IEEEkeywords}
Android Malware Detection, Machine Learning, Evasion Attacks, Adversarial Purification, Denoising Autoencoder.
\end{IEEEkeywords}}

\maketitle

\IEEEdisplaynontitleabstractindextext

% \fi
\IEEEpeerreviewmaketitle

\IEEEraisesectionheading{\section{Introduction}\label{section1}}
\IEEEPARstart{D}{ue} to the popularity of the Android operating system, it has become the primary victim of malware attacks. In 2021, Zimperium reported that 2 billion new malware emerged in the wild~\cite{2022GMTR}, and Kaspersky detected 1,661,743 mobile malware or unwanted software installers in 2022~\cite{mmt2022}. As a result, the prevalence of Android malware has grown exponentially in recent years, posing a significant threat to the security and privacy of mobile users worldwide.

The magnitude of this threat has spurred the use of Machine Learning (ML) techniques, particularly Deep Learning (DL), to automate \emph{Android malware detection}. Empirical evidence has shown that these approaches offer advanced performance in detecting malware (see, e.g.,~\cite{zhu2021sedmdroid,xu2022sdac,qiu2022cyber,zhu2022hybrid,fang2023comprehensive}), making them a promising avenue for mitigating this security concern. 

Despite these advances in detection capabilities, ML-based detectors are vulnerable to adversarial examples, which are created by modifying non-functional instructions in executable programs of existing malware~\cite{pierazzi2020intriguing,li2023black}. Adversarial examples can enable a range of attacks, including \emph{evasion attacks}~\cite{demontis2019yes,chen2020android,li2020adversarial}, \emph{poisoning attacks}~\cite{li2022backdoor,severi2021explanation,suciu2018does}, or a combination of both~\cite{demontis2019adversarial}. In this study, we specifically narrow our focus to evasion attacks, which are designed to deceive ML-based detection during the testing phase.

So far, \emph{adversarial training}-based methods have shown great potential to safeguard ML models from evasion attacks~\cite{zhou2024mtdroid,qiao2023adversarial,li2023pad}. By augmenting the training dataset with generated adversarial samples, adversarial training can increase the robustness of the trained model in future use. However, these methods still have certain disadvantages, including high computational costs~\cite{jia2022boosting} and a significant sacrifice in accuracy on clean data~\cite{lau2023interpolated}. Furthermore, its effectiveness is strongly influenced by the similarity between the adversarial examples employed during the training and testing phases~\cite{li2021robust}. This may lead to overfitting of the model to specific perturbations, thereby negatively impacting its ability to generalize and detect unseen attacks~\cite{cheng2026constraint}.

Another defense technique, known as \emph{adversarial purification}~\cite{naseer2020self,yoon2021adversarial,croce2022evaluating}, aims to remove potential perturbations from input samples, resulting in \emph{purified} samples that can be correctly classified by the target classifier. The purification model is usually trained independently of the classification model and does not necessarily require class labels~\cite{theagarajan2022privacy}. As a result, it can mitigate unseen threats in a \emph{plug-and-play} manner without re-training the target classifier~\cite{nie2022diffusion}, leading to less training overhead and more flexible employment. Nevertheless, existing purification solutions for image classification are not easily applicable to Android malware detection due to significant differences~\cite{xu2023ofei} as: (i) The feature space of Android applications is not only discrete but also high-dimensional, making noise removal fundamentally different from denoising continuous pixel values in the image domain. (ii) Evasion attacks in this domain are far more diverse than simple perturbations, including complex structural and semantic manipulations that require more than a simple reconstruction objective to defend against. (iii) A practical defense must maintain high accuracy on benign samples, as false positives can render security products unusable. However, this constraint is less stringent in other domains.

\begin{figure}[t]
  \centering
  \includegraphics[width=0.43\textwidth]{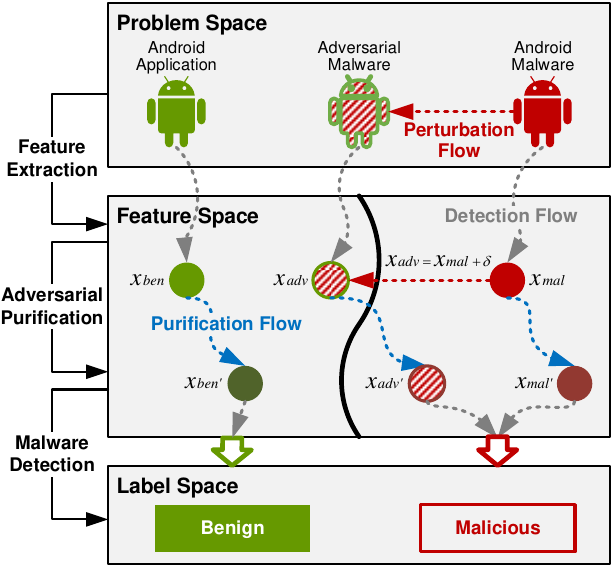}
  \caption{Illustration of MalPurifier pre-processing samples via adversarial purification, before feeding them into the malware detector. This method projects the various perturbed samples back to their original forms, while reserving the feature representation of clean data, successfully striking a balance between robustness and accuracy without necessitating any changes to the architecture or parameters of the detection model.}
  \label{illustration}
\end{figure}

To overcome these challenges, we propose MalPurifier, a novel adversarial purification framework specifically engineered for the Android malware ecosystem, as illustrated in Fig.~\ref{illustration}. MalPurifier introduces three key innovations as follows: (i) To defend against a wide range of unforeseen attacks, we introduce a mechanism that generates adversarial malware with progressively increasing perturbation strengths, from zero perturbation to worst-case manipulation. This exposes the purifier to a broad spectrum rather than a fixed distribution, enabling to learn a more generalizable representation of maliciousness. (ii) Motivated by the asymmetric threat model in which adversaries primarily target malware rather than benign applications, we propose a novel strategy that injects ``\emph{protective noises}" into benign samples. This teaches the purifier to preserve the integrity of legitimate applications and avoid costly false positives. (iii) To bridge the gap between purification quality and classification accuracy, we design a \emph{Denoising AutoEncoder} (DAE) with a tailored loss function that jointly optimizes both both reconstruction fidelity and feature-space alignment with the downstream detector, going beyond the plain reconstruction loss used in prior works. As a result, MalPurifier operates as a \emph{non-intrusive}, \emph{computationally efficient}, and \emph{easily scalable} plug-and-play module that significantly enhances the robustness of existing malware detectors. The main contributions of this paper can be summarized as follows:

\begin{itemize}[leftmargin=*]
  \item \textbf{Novel purification framework for Android evasion attacks.} We propose \emph{MalPurifier}, a robust purification framework tailored for the unique challenges of the Android ecosystem. Instead of a generic application of autoencoders, our method introduces novel mechanisms specifically designed to handle the discrete feature space and diverse attack vectors inherent to Android malware.
  \item \textbf{Trade-off between robustness and detection accuracy.} Our proposed mechanism injects a diversified range of perturbations into malware samples, ranging from no injection to worst-case perturbations, to enhance robustness against different attacks, especially unforeseen attacks. We also implement a novel protective noise injection strategy to prevent false positives on benign samples.
  \item \textbf{Accurate sample recovery via denoising autoencoder.} We establish a DAE-based purification model for purifying adversarially perturbed samples, which is independent of the label space and the detection model. Especially, we incorporate reconstruction loss and prediction loss to help the model handle complex and noisy data, leading to better feature representation and sample recovery.
  \item \textbf{Experimental validation on public datasets.} We compare MalPurifier with the state-of-the-art (SOTA) methods via \emph{Drebin}~\cite{arp2014drebin} and \emph{Androzoo}~\cite{allix2016androzoo} datasets. Experimental results show that MalPurifier significantly outperforms other defenses against a comprehensive set of perturbation-based and structure-based evasion attacks, with less overhead required. Finally, we release the source code at \url{https://github.com/SEU-ProactiveSecurity-Group/MalPurifier}.
\end{itemize}

The remainder of this paper is organized as follows. Section~\ref{section2} reviews some preliminaries and Section~\ref{section3} presents the problem formulation. Section~\ref{section4} elaborates the methodology, and the experimental results are presented in Section~\ref{section5}. We explore the limitations of our work and open challenges in Section~\ref{section6}, and discuss related work in Section~\ref{section7}. Finally, conclusion and future work are summarized in Section~\ref{section8}.

\section{Preliminaries}\label{section2}
This section provides the necessary background for understanding our approach. We first review ML-based Android malware detection in Section~\ref{section2.1}, then examine various evasion attack methods in Section~\ref{section2.2}, and finally introduce the concept of adversarial purification in Section~\ref{section2.3}.

\subsection{ML-based Malware Detection}\label{section2.1}
The ML-based Android malware detection can be briefly described as follows. Formally, let $\mathcal{Z}$ be the problem space, and $z \in \mathcal{Z}$ be an Android application sample. In the context of machine learning, there will be a feature extraction function $\phi:\mathcal{Z}\rightarrow \mathcal{X}$ which maps the problem space into the feature space, where $\mathcal{X}\subset \mathbb{R}^d$ is a $d$-dimensional feature space. 

The Android malware detection can be usually treated as a binary classification, thus, let $f:\mathcal{Z}\rightarrow \mathcal{Y}$ be the malware detector that maps the problem space to the label space $\mathcal{Y}=\{0,1\}$, where "0" (or "1") means that corresponding example is benign (or malicious), respectively. Additionally, let the malware detector use an ML model $\varphi_{\theta}:\mathcal{X}\rightarrow \mathcal{Y}$, where $\theta$ represents the model's parameters. Therefore, we can conclude that $f(\cdot)=\varphi_{\theta}(\phi(\cdot))$.

Given a sample-label pair $(z,y)$ and ML-based malware detector $f$, we then have $x=\phi(z)$. We can easily achieve the prediction $f(z)$ and compare it with the ground-truth label $y$ to analyze the accuracy. To improve the detection accuracy, the main task of Android malware detection is to achieve the optimal parameters as follows.
\begin{equation}\label{eq1}
  \theta^* \in \arg \min_{\theta} \mathbb{E}_{(z,y)\in \mathcal{D}}[\mathcal{L}(\theta,x,y)],
\end{equation}

\noindent where $\mathcal{L}(\theta,x,y)$ is the loss function for the model $\varphi_{\theta}$, and $\mathcal{D}$ represents data distribution of training examples.

\subsection{Evasion Attacks}\label{section2.2}
\subsubsection{Attack Principle}
According to Ref.~\cite{pierazzi2020intriguing}, evasion attack can be categories into two types: \emph{problem-space} attacks and \emph{feature-space} attacks. In the problem-space attack, the adversary perturbs a malware sample from $z$ to $z'$ to evade the detector $f$. Accordingly, they can be mapped into the feature space with $x=\phi(z)$ and $x'=\phi(z')$. Formally, given a feature-label pair $(x,y)$ of a malware sample and an adversarial manipulation $\delta$, the evasion attack can be written as
\begin{equation}
  \varphi_{\theta}(x')=\varphi_{\theta}(x+\delta)=0, \quad  \mathrm{s.t.} (x' \in \mathcal{X}) \land (x' \in [\check{u},\hat{u}]),
\end{equation}

\noindent where $x'$ is the perturbed feature representation. Recent studies have suggested that it obeys a box constraint~\cite{demontis2019adversarial}, such that $x' \in [\check{u},\hat{u}]$, where $\check{u}$ and $\hat{u}$ denote the lower and upper boundaries in the feature space, respectively.

To establish the inverse mapping from the feature space to the problem space, we adopt the methodology proposed in Ref.~\cite{vsrndic2014practical}. This approach facilitates the design of attack tactics while maintaining the effectiveness of the attack. By utilizing an approximate inverse function $\tilde{\phi}^{-1}$, we can directly map the perturbation vector $\delta$ to the problem space. 

\subsubsection{Attack Methods}
\noindent \textbf{Obfuscation Attacks.}
This kind of attacks suggests malware authors leveraging obfuscation technology to camouflage malicious functionality~\cite{jeon2017avpass,demontis2019yes}. Typically, adversaries exploit certain techniques (e.g., encryption, renaming, etc.) to produce malware variants that can deceive detection. Note that this attack does not require knowledge of the target classifier, making it a zero-query black-box attack that can be directly performed on the problem space. 

\noindent \textbf{Gradient-based Attacks.}
These attacks apply small perturbations in the direction of gradients to produce adversarial malware samples. For example, Projected Gradient Descent (PGD) attack~\cite{madry2018towards} initializes the perturbation with a zero vector and perturbs it via an iterative process, such that
\begin{equation}
  \delta^{t+1} = \Pi_{[\check{u}-x,\hat{u}-x]}\Big(\delta^{t}+\lambda\nabla_{\delta}\mathcal{L}(\theta,x+\delta^{t},y)\Big),
\end{equation}

\noindent where $t$ is the iteration, $\lambda>0$ is the step size, $\Pi_{[\check{u}-x,\hat{u}-x]}$ is the projection operator that keeps $\delta^{t+1}$ within a set of range $[\check{u}-x,\hat{u}-x]$, and $\nabla_{\delta}$ indicates the gradient 
of the loss function $\mathcal{L}$ with respect to $\delta$. Due to the small magnitudes of gradients in practical scenarios, researchers have been motivated to normalize the gradients in a direction of interest, such as the $\ell_1$, $\ell_2$, or $\ell_\infty$ norm~\cite{zhang2019interpreting}. 

Furthermore, this study incorporates several other algorithms to perform gradient-based attacks, including Bit Coordinate Ascent (BCA)~\cite{al2018adversarial}, Fast Gradient Sign Method (FGSM)~\cite{qiao2023adversarial}, and Grosse~\cite{hu2022generating}.

\noindent \textbf{Gradient-free Attacks.}
These attacks are permitted to get access to a surrogate dataset and wage evasion attacks via perturbations. The salt and pepper noises attack~\cite{li2020adversarial} involves manipulating malware samples by randomly replacing feature values with either the maximum or minimum intensity values, resembling the spread of salt and pepper particles. This study also investigates the use of pointwise attacks~\cite{vo2021query}, in which the adversary first adds noise perturbation and then modifies features to generate an adversarial sample with the least perturbation.

\noindent \textbf{Ensemble Attacks.}
These attacks provide attackers with the capability to compromise the victim via a combination of multiple attack methods and manipulations. For instance, Li et al. \cite{li2023pad} proposed a series of ensemble-based attacks, including the "Max" strategy enabled Mixture of Attacks (MaxMA), iterative MaxMA (iMaxMA), and Stepwise Mixture of Attacks (StepwiseMA), which effectively enhance the attack performance. Additionally, Croce and Hein \cite{croce2020reliable} combined powerful attacks to create an ensemble attack namely AutoAttack, which demonstrates strong generalization across different models.

\subsection{Adversarial Purification}\label{section2.3}
To counter these diverse evasion attacks, adversarial purification~\cite{nguyen2025pbp} has emerged as a promising defense strategy. The fundamental concept behind it is to preprocess the input data directly, preventing any embedded adversarial components from feeding into the target model, so that the influence of attacks can be mitigated. These methods are widely regarded as model-agnostic and highly efficient, making them easy to train and utilize while demonstrating strong generalization capabilities.

Let $g$ be the adversarial purifier that uses a generative model $\psi_\vartheta$ with $g(\cdot)=\psi_\vartheta(\phi(\cdot))$ to learn the data distribution closer to the training distribution and restore an adversarial example to its corresponding clean example, where $\vartheta$ represents its parameters. Given $x=\phi(z)$ and $x'=\phi(z')$, thus, the training objective of purification is then
\begin{equation}\label{eq4}
  \begin{aligned}
  \vartheta^* \in \arg & \min_{\vartheta} \mathbb{E}_{(z,y)\in \mathcal{D}}[\mathcal{J}(\vartheta,x',x)],\\
  \mathrm{s.t.} (x' \in \mathcal{X})&\land(x' \in [\check{u},\hat{u}])\land(\psi_\vartheta(x') \in \mathcal{X}),
  \end{aligned}
\end{equation}

\noindent where $\mathcal{J}(\vartheta,x',x)$ represents the loss function for the learning model $\psi_\vartheta$, $x$ denotes the original feature, and $x'=x+\delta$ is the perturbed feature representation. This training procedure only focuses on the differences of the representation between the sample after purification and its original version, thus, we can clearly conclude that the purification model is trained independently of the class label.

Unlike adversarial example attacks in the image domain that perturb images with inconspicuous noises, adversaries in this field specifically employ discrete manipulations on malware samples to evade detection. These adversarial examples closely resemble benign data in their feature representation, posing a significant challenge for the purification model. For example, clean data might be mistaken for adversarial examples, resulting in incorrect restoration into samples with malicious feature representation. As a consequence, the accuracy of clean data may experience a substantial decrease. Thereby, it remains a question of how to effectively enhance the trade-off between robustness and accuracy of adversarial purification.

\section{Problem Formulation}\label{section3}
Here we introduce the threats considered in our work, and propose a defense formulation to guide the defense design.

\subsection{Threat Model}
We consider the threat model in terms of assumptions regarding the attacker's capabilities and knowledge of the target system, specified by three attack scenarios as follows.

\subsubsection{Black-Box Attacks}
In this attack scenario, the attacker has no knowledge about either the malware detector $f$ or the purifier $g$. Given a malware example $z$, the attacker attempts to perturb it from $z$ to $z'$, resulting in a feature space transformation from $x=\phi(z)$ to $x'=\phi(z')$. The attacker's goal is to make the prediction of the target system incorrect, such that
\begin{equation}
  \varphi_{\theta}(\psi_\vartheta(x'))=0,\quad \mathrm{s.t.} (x' \in \mathcal{X})\land(x' \in [\check{u},\hat{u}]),
\end{equation}

\noindent where the range set of $[\check{u},\hat{u}]$ ensures the feasibility of manipulations and the persistence of malicious functions.

\subsubsection{Grey-Box Attacks}
In the case of grey-box attacks, since the adversary is aware of the characteristics of the malware classifier $f$ but is oblivious to the structure of the purifier $g$, we also refer to this attack pattern as \emph{oblivious attacks}. Hence, attack strategy here focuses on deceiving the unsecured classifier without considering the purifier. Formally, given a malware example $z$ with its feature representation $x=\phi(z)$ and label $y=1$, the attacker will modify it to obtain the adversarial feature representation $x'$ that can evade detection, by solving
\begin{equation}
  \max_{x' \in [\check{u},\hat{u}]} \mathcal{L}(\theta,x',1), \quad \mathrm{s.t.} x' \in \mathcal{X},
\end{equation}

\noindent where we substitute $\varphi_\theta(x')=0$ with maximizing $\mathcal{L}(\theta,x',1)$ owing to the non-differentiability of $\phi(\cdot)$.

\subsubsection{White-Box Attacks}
In the white-box attack setting, the attacker is granted complete knowledge of both the target model $f$ and the purifier's architecture $g$, thus constituting what is formally known as \emph{adaptive attacks}. In this scenario, the adversary aims to craft more sophisticated attacks that mislead both the malware detector and adversarial purifier simultaneously. Therefore, given a malware feature-label pair $(x,y)$, the attacker needs to perturb $x$ into $x'$ by solving
\begin{equation}\label{eq7}
  \max_{x' \in [\check{u},\hat{u}]} \mathcal{L}(\theta,\psi_\vartheta(x'),1), \quad \mathrm{s.t.} (x' \in \mathcal{X})\land(\psi_\vartheta(x') \in \mathcal{X}),
\end{equation}
\noindent where $\psi_\vartheta(x')$ denotes the purified sample obtained by applying the adversarial purifier $g$ to the input $x'$.

\subsection{Defense Formulation}
As aforementioned, MalPurifier is rooted in adversarial purification, aiming to eliminate potential adversarial manipulations before detection. Thereby, we propose incorporating the detector $f$ with the adversarial purifier $g(\cdot)=\psi_\vartheta(\phi(\cdot))$. To develop the framework of MalPurifier, we need to train the detection model $\varphi_{\theta}$ according to Eq.~(\ref{eq1}) and build the purification model $\psi_{\vartheta}$ based on Eq.~(\ref{eq4}), respectively. To this end, given a feature-label pair $(z,y)$ with possible adversarial perturbations $\delta$ in the feature space, the desired parameters $\theta^*$ and $\vartheta^*$ of MalPurifier can be derived by solving the following problem
\begin{equation}
  \begin{aligned}
  \varphi_{\theta^*}(\psi_{\vartheta^*}(x&+\delta))=y,\\
  \mathrm{s.t.} (x \in \mathcal{X})\land(y \in &\mathcal{Y})\land(x+\delta \in [\check{u},\hat{u}]),
  \end{aligned}
\end{equation}

\noindent when $\delta=0$, it indicates that no adversarial manipulations have been applied to this sample, rendering it clean. The above formulation points out two tasks as follows.

\begin{itemize}[leftmargin=*]
  \item \textbf{High-accuracy Android malware detection.} Developing a malware detection model that can classify clean samples into benign or malicious with high accuracy, that is, $\varphi_{\theta}(x)=y$, in which $x$ can be the feature representation of a normal Android application or a malware sample, and thus $y=0$ or $1$. This model can be easily obtained by utilizing some ML algorithms (e.g., Deep Neural Network (DNN)), which have shown promising results with 99\% accuracy in their laboratory settings~\cite{kim2019multimodal,chen2020android,zhu2022hybrid}.
  \item \textbf{Effective adversarial manipulation elimination.} This formulation highlights the importance of effectively integrating adversarial purification with the pre-trained detector, as summarized into two aspects. (i) Given the feature representation of a malware sample $x$ with its adversarial version $x'$, it is crucial to minimize the impact of evasion attacks by achieving $\psi_{\vartheta}(x')\thickapprox x$. This enables us to accurately identify it through the malware detector, indicated by $\varphi_{\theta}(\psi_{\vartheta}(x'))=1$. (ii) When dealing with a clean feature-label pair $(x,y)$, it is essential to preserve the accuracy on it. This entails ensuring that the prediction results of clean data remain unaffected, represented by $\varphi_{\theta}(\psi_{\vartheta}(x))=y$.
  \end{itemize}

\section{The MalPurifier Approach}\label{section4}
In this section, we present the MalPurifier approach in detail. We first provide an overview of the system architecture, including both the training and inference processes. Next, we describe the mechanisms for generating diversified adversarial perturbations and injecting protective noise. These mechanisms are crucial for creating robust training data for the purification model. Finally, we introduce the label-independent DAE-based purification model, which is responsible for producing purified samples and improving detection accuracy.

\begin{figure*}[t]
  \centering
  \includegraphics[width=0.95\textwidth]{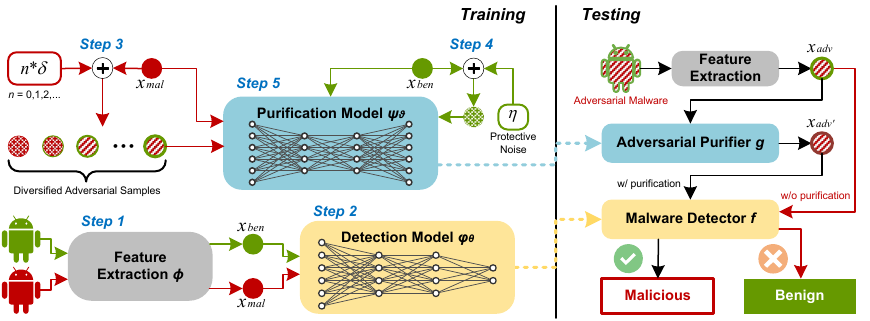}
  \caption{Overview of MalPurifier architecture. In the training phase, feature vectors are extracted from Android apps in Step 1. Then, a detection model is constructed using features from both benign and malicious apps in Step 2. In Step 3, diversified adversarial perturbations are applied to malware samples in the feature space, while protective noises are introduced into benign samples in Step 4. Finally, in Step 5, the purification model is built using these variant samples along with their corresponding original versions. In the testing phase, a sample undergoes sequential processing by the purifier and detector, ensuring that adversarial malware cannot escape detection.}
  \label{overview}
\end{figure*}

\subsection{Architecture Overview}\label{section4.1}
Figure~\ref{overview} provides an overview of the MalPurifier architecture. As we can see, it is composed of three main modules: (i) a feature extractor $\phi(\cdot)$ that maps an Android application into a feature vector, (ii) an adversarial purifier $g$ that processes samples via a DAE model $\psi_{\vartheta}$, and (iii) a malware detector $f$ that uses a DNN model $\varphi_{\theta}$ for detection.

During the training phase, we first extract features from batches of clean (or natural) data. These samples include both benign and malicious examples without manipulation. These features are then fed into the model (e.g., DNN), which iteratively updates its parameters to minimize the loss function. Once the DNN model is effectively trained, it exhibits exceptional classification accuracy when presented with clean inputs. Notably, we do not retrain the DNN with labeled adversarial data. This key aspect distinguishes our approach from traditional adversarial training methods.

In contrast, MalPurifier incorporates a purification model (e.g., DAE). This model is specifically designed to learn compact representations of input data and reconstruct the original (clean) data from its noisy or perturbed versions. The effectiveness of the DAE largely depends on the quality and diversity of the training data, particularly the types and levels of noise introduced during training.

Unlike prior purification methods that train solely on a specific attack type and thus fail to generalize across diverse attack types, we first propose a diversified adversarial perturbation mechanism (see Section~\ref{section4.2}). Rather than assuming a fixed perturbation distribution, this mechanism generates adversarial malware with progressively increasing perturbation strengths,from zero perturbation to worst-case manipulation, exposing the DAE to a broad spectrum of adversarial behaviors and enabling it to generalize to previously unseen attacks.

Furthermore, a critical yet often overlooked challenge in the malware domain is the risk of over-purification on benign samples, which can inflate false positive rates. To explicitly address this, we propose a noise injection strategy for benign samples (see Section~\ref{section4.3}). This strategy is specifically motivated by the asymmetric threat model in malware detection. Since adversaries target malware rather than benign apps, we inject controlled random noise into benign samples during training, explicitly teaching the purifier to preserve the recognition of legitimate applications, which is absent in prior purification approaches.

In addition, beyond the plain reconstruction loss used in conventional methods that optimizes only for feature-level fidelity, we incorporate both reconstruction loss and prediction loss into the DAE's objective function. This dual-objective design directly bridges the gap between purification quality and classification accuracy, as the prediction loss encourages the purified output to align with the original sample in the detector's internal feature space. The model is trained using a combination of adversarially perturbed malware samples, noisy benign samples, and their corresponding clean counterparts. Importantly, the DAE is trained independently of class labels, which enhances its generalization capability. Further details on the DAE training process are provided in Section~\ref{section4.4}.

During the testing phase, an input sample is sequentially processed by MalPurifier's feature extraction, purification, and detection modules to yield a prediction result, as illustrated in the right part of Fig.~\ref{overview}.

\subsection{Diversified Adversarial Perturbation}\label{section4.2}
To address the challenge of diverse attack vectors in the malware domain, a simple perturbation strategy (e.g., adversarial training) is insufficient. To this end, we propose a diversified adversarial perturbation mechanism designed to enhance the generalizability of the purification model. 

By exposing the model to a wide variety of perturbations, it enables the purifier to defend not only against known attack types but also to effectively mitigate previously unseen or unknown attacks. Therefore, our approach aims to maximize the difference in feature space between the original malware sample $x$ and its perturbed counterpart $x'$. This objective can be formally expressed as
\begin{equation}\label{eq9}
  \Delta(x,x')=d(\mathcal{F}_\theta(x)|n,\mathcal{F}\theta(x')|n), \ \ \mathrm{s.t.} x' \in [\check{u},\hat{u}], 
\end{equation} 
\noindent where $\mathcal{F}\theta(x)|_n$ is the internal feature representation of $x$ at the $n$th layer of the malware detector $f$, and $d(\cdot)$ denotes the distance metric used to quantify the difference between the original and perturbed features. In our implementation, we use \emph{Mean Square Error} (MSE) as the distance metric.

\begin{algorithm}[t]\label{algorithm1}
  \caption{Diversified Adversarial Perturbation}
  \LinesNumbered
  \KwIn{Training dataset $(\mathcal{X},\mathcal{Y})$, number of batches $N$, number of iterations $T$, step size $s$, and random transformation function $\mathcal{R}$\;}
  \KwOut{Generated adversarial subset $\mathcal{X}_{adv}$\;}    
  \For{$i=1$ to $N$}{
    Sample a batch of $(x_i,y_i=1)$ from $(\mathcal{X},\mathcal{Y})$\;
    Adversarial depth $k_i=s*(i-1)$\;    
    \If{$k_i=0$}{
      $x_i' \leftarrow x_i$; \Comment{First batch without perturbation}\\
    }
    \Else{
      $x_{i,1}' =\mathcal{R}(x_i)$; \Comment{Create a random initial point}\\
      \For{$t=1$ to $T$}{
        Compute $\Delta$ between $x_i$ and $x_{i,t}'$ via Eq.(\ref{eq9})\;
        Compute gradients $g_t=\nabla_{x_i}\Delta(x_i,x_{i,t}')$\;
        Generate adversarial samples by $x_{i,t+1}'=x_{i,t}'+k_i*g_t$\;
      }
      $x_i' \leftarrow x_{i,T}'$; \Comment{Other batches with perturbation}\\
    }
  }
  Return $\mathcal{X}_{adv}=\{x_1',x_2',...,x_N'\}$
\end{algorithm}

Following prior work~\cite{pierazzi2020intriguing,li2023pad}, perturbations are constrained in a way to preserve the malicious functionality of the original malware. Specifically, for manifest-derived features, this corresponds to injecting non-functional operations, such as unused hardware declarations, unrequested permissions, or redundant implicit intent registrations. These additions would not alter the app's runtime behavior. For dexcode-derived features, perturbations consist of inserting Application Programming Interface (API) calls (e.g., dynamic code loading, or low-level command execution) within unreachable code segments, thereby introducing new feature activations without affecting the program's control flow. Note that we exclude features that can be trivially renamed or modified (e.g., package name), as they do not reliably reflect malicious behavior.

The process for generating diversified adversarial examples is outlined in Algorithm~\ref{algorithm1}. The main steps are as follows: (i) We begin by sampling a batch of malware examples $(x_i, y_i=1)_{i=1}^N$ in Line 2. (ii) For each batch, we set the adversarial depth in proportion to the batch index in Line 3, so that the perturbation level is gradually increased across batches to cover different attack intensities. (iii) For the first batch, no perturbations are applied in Line 5. (iv) For subsequent batches, we initialize $x_i$ with a random transformation in Line 7. We then compute the gradient $g_t$ based on the feature difference in Line 10 and iteratively update the sample according to the preset adversarial depth in Line 11. (v) Specifically, for Drebin features, we additionally utilize clamping and randomized rounding~\cite{al2018adversarial}, guaranteeing that each generated adversarial sample is projected back into the binary feature space. For continuous-valued features, this projection step is omitted, as the gradient-based perturbations naturally reside in the continuous feature space. (vi) After $T$ iterations, we obtain the final adversarial samples for the batch in Line 12. (vii) By repeating these steps for all batches, we obtain a comprehensive set of adversarial examples with varying perturbation strengths, ensuring both diversity and generalizability for robust purification.

\subsection{Protective Noise Injection}\label{section4.3}
While the diversified perturbation strategies described in Section~\ref{section4.2} can significantly enhance the generalizability of the purification model against various evasion attacks, they may also introduce a new critical challenge, such that the risk of over-purification on clean data. Excessive purification can inadvertently corrupt benign samples and increase the false positive rate. Moreover, in practice, adversaries rarely attempt to modify benign samples to mimic malware. Therefore, directly applying adversarial perturbations to benign data is both unrealistic and unnecessary.

\begin{algorithm}[t]\label{algorithm2}
  \caption{Protective Noise Injection}
  \LinesNumbered
  \KwIn{Training dataset $(\mathcal{X},\mathcal{Y})$, number of batches $N$, number of iterations $T$, and noise level $\eta$\;}
  \KwOut{Processed benign subset $\mathcal{X}_{ben}$\;}    
  \For{$i=1$ to $N$}{
    Sample a batch of $(x_i,y_i=0)$ from $(\mathcal{X},\mathcal{Y})$\;
    Obtain its batch size $b_i$ and length $l_i$\;
    Generate random mask $m=rand(b_i,l_i)<\eta$\;
    Flip feature values via $x_i[m]=1-x_i[m]$\;
  }
  Return $\mathcal{X}_{ben}=\{x_1,x_2,...,x_N\}$
\end{algorithm}

To tackle this, we introduce a novel protective noise injection mechanism that operates exclusively during the training phase with controlled random noise. The goal is \textit{not} to add noise to benign samples at inference time, but rather to expose the purification model to a diverse distribution of benign inputs during training, so that it learns to preserve the recognizability of legitimate applications. Specifically, we define a threshold parameter $\eta \in [0,1]$ to control the extent of noise injection ($\eta=0$ means no noise added while $\eta=1$ indicates that all features are perturbed). By tuning $\eta$, we can balance the trade-off between robustness and accuracy, ensuring that the purification model maintains high detection performance on clean data.

Algorithm~\ref{algorithm2} presents the detailed steps of this mechanism: (i) For each batch in the training set, we first select all benign samples $(x_i, y_i=0)$ in Line 2. (ii) We then determine the batch size $b_i$ and the feature length $l_i$ for subsequent processing in Line 3. (iii) Next, we randomly generate a binary mask based on the preset noise level $\eta$ to identify which feature positions will be altered in Line 4. (iv) The selected feature values are then perturbed according to the generated mask in Line 5. (v) By repeating these steps for all batches, we obtain an augmented set of benign samples with protective noise in Line 6. These modified benign samples are subsequently used to train the purification model.

\subsection{DAE-based Purification Model}\label{section4.4}
Building on the mechanisms described in Sections~\ref{section4.2} and~\ref{section4.3}, we construct a comprehensive training dataset by combining diversified adversarial malware samples, benign samples with protective noise, and their original clean counterparts. Therefore, in this section, we present the technical details of our DAE-based purification model and explain how it effectively restores perturbed malware samples while preserving the detection accuracy for clean data.

While various generative models can be used for purification, we specifically adopt a DAE for the Android malware detection, which is motivated by following reasons: (i) Compared to classical methods like \emph{Principal Component Analysis} (PCA), DAE is particularly effective at removing diverse and structured noise from discrete, high-dimensional feature spaces, which matches the nature of Android application features, whether binary or continuous-valued. (ii) Unlike generative models such as \emph{Variational Autoencoder} (VAE) and \emph{Generative Adversarial Network} (GAN), DAE does not require adversarial objectives or complex regularization, making it more scalable and robust in practice. (iii) Compared to \emph{Diffusion Model} (DM), DAE is easier to train and model-agnostic, allowing it to be integrated as a plug-and-play module for downstream detectors.

As illustrated in Fig.~\ref{overview}, the purification model is trained with several types of input: perturbed malware samples (generated via Algorithm~\ref{algorithm1}), benign samples modified by protective noise (via Algorithm~\ref{algorithm2}), as well as their original, unperturbed forms. During training, these data are passed through the network, and the parameters $\vartheta$ are optimized to minimize a customized loss function as described below.

\noindent \textbf{Reconstruction Loss.} A standard loss function for DAE is the MSE loss, which measures the average squared difference between the reconstructed output and the target data. To train a robust purification model, we minimize the discrepancy between the reconstructed output and the original data (i.e., reconstruction loss) as follows.
\begin{equation}
  \mathcal{L}_{rec}=d(x,\psi_\vartheta(x')), \ \ \mathrm{s.t.} x' \in \mathcal{X}_{adv}\cup \mathcal{X}_{ben},
\end{equation}

\noindent where $x$ and $\psi_\vartheta(x')$ denote the original data and the purified data, respectively.

\noindent \textbf{Prediction Loss.} Since the ultimate goal of adversarial purification is to improve the classification performance of the downstream malware detector, we introduce a prediction loss to further optimize the purification model as follows.
\begin{equation}
  \mathcal{L}_{pre}=\Delta(x,\psi_\vartheta(x'))=d(\mathcal{F}_\theta(x)|_n,\mathcal{F}_\theta(\psi_\vartheta(x'))|_n),
\end{equation}
\noindent where $\Delta$ is formally defined in Eq.~(\ref{eq9}), and $d(\cdot)$ is again the MSE. This loss encourages the purified data to be close to the original data in the internal feature space of the malware detector, leading to more accurate predictions.

The overall loss function is a weighted combination of the reconstruction loss and the prediction loss as follows.
\begin{equation}
  \mathcal{L}_{\psi_\vartheta}=\alpha\mathcal{L}_{rec}+\beta\mathcal{L}_{pre},\ \ \mathrm{s.t.} \alpha,\beta \in [0,1],
\end{equation}

\noindent where $\alpha$ and $\beta$ are the weights of two loss terms, and we have $\alpha+\beta=1$. The parameters $\vartheta$ are optimized by minimizing this combined loss during training.

\section{Experiments and Evaluation}\label{section5}
In this section, we conduct extensive experiments to validate the soundness of MalPurifier by answering the following Research Questions (RQs):
\begin{itemize}[leftmargin=*]
  \item \textbf{RQ1: Effectiveness and cost without attacks.} How is the effectiveness and overhead of MalPurifier when there is no attack?
  \item \textbf{RQ2: Robustness against black-box attacks.} How is the robustness of MalPurifier against black-box attacks?
  \item \textbf{RQ3: Robustness against grey-box attacks.} How robust is MalPurifier against grey-box attacks where the attacker is unaware of the additional defense mechanism (e.g., the adversarial purifier $g$)?
  \item \textbf{RQ4: Robustness against white-box attacks.} How robust is MalPurifier against white-box attacks in which the adversary has full knowledge of all defense mechanisms?
  \item \textbf{RQ5: Interpretability of the purification process.} How does the purification process affect feature representations, and which features are most relevant to the correction of adversarial perturbations?
  \item \textbf{RQ6: Advantage and transferability of the purifier.} Does the DAE model in MalPurifier outperform alternative purification techniques, and can it be flexibly transferred to enhance other types of detectors against evasion attacks?
  \item \textbf{RQ7: Generalizability against structural attacks.} How does MalPurifier perform when applied to graph-based feature representations against attacks that target the structural or semantic properties of Android applications?
\end{itemize}

\noindent \textbf{Datasets.} To evaluate MalPurifier under both a standardized benchmark setting and a temporally diverse threat environment, our experiments utilize two popular Android malware datasets: \emph{Drebin}~\cite{arp2014drebin} and \emph{Androzoo}~\cite{allix2016androzoo}. The Drebin dataset\footnote{https://www.sec.cs.tu-bs.de/~danarp/drebin} consists of 5,560 malicious samples and SHA256 values of 123,453 benign applications, which were collected before 2013. For evaluation purposes, we downloaded 47,770 benign APKs from various markets (e.g., Google Play Store, AppChina, Anzhi). To obtain more recent files, we collected 170,851 APKs from the Androzoo dataset\footnote{https://androzoo.uni.lu}, specifically those attached with dates falling between January 1st and December 31st, 2021. We submitted these APKs to the \emph{VirusTotal}\footnote{https://www.virustotal.com/gui/home/upload} service, labeling a sample as malicious if at least five anti-virus scanners raised alarms, and considering it benign if no scanner detected it. We randomly selected 10,987 benign examples and 10,998 malicious examples from Androzoo for our experiments. Note that each dataset was randomly split into three distinct sets for training (60\%), validation (20\%), and testing (20\%).

\noindent \textbf{Feature extraction.} Drebin~\cite{arp2014drebin} analyzes a set of APKs and constructs a suitable feature space. We here utilize the \emph{Androguard}\footnote{https://github.com/androguard/androguard} tool to perform a static analysis and extract the Drebin features, extracted from the \texttt{manifest} (e.g., hardware, requested permissions, intents), and the disassembled \texttt{dexcode} (e.g., restricted API calls, suspicious API calls). The APK is mapped into the feature space as a binary feature vector, in which we can have 0 or 1 along each dimension, indicating the presence or absence of the corresponding feature. Following prior work~\cite{li2023pad}, we exclude certain features that can be easily renamed or modified (e.g., package name) and retain the most frequent 10,000 ones in this study. 

Note that the above Drebin features are used as the default feature representation for RQ1–RQ6. To further evaluate MalPurifier's generalizability to graph structures, we additionally conduct experiments using API call sequences extracted from the applications' Control Flow Graphs (CFGs) in RQ7. Specifically, following the MaMaDroid~\cite{onwuzurike2019mamadroid} methodology, we extract the call graph of each application via static analysis, abstract the API calls to their package names, and build a Markov chain to model the transitions between packages. The resulting transition probability matrix is used as the feature representation, and PCA with 10 components is applied to reduce dimensionality. Under this setting, all defense methods are retrained from scratch using the new feature representation to ensure a fair comparison.

\begin{table*}[t]
  \caption{Performance on clean data of Drebin and Androzoo datasets, where effectiveness metrics are in percentage (\%) and training time is in seconds (s).}
  \centering
  \label{table1}
  \begin{tabular}{@{}l|cccccc|cccccc@{}}
  \toprule
  \multirow{2}{*}{Defense} & \multicolumn{6}{c|}{Drebin} & \multicolumn{6}{c}{Androzoo} \\ \cmidrule(l){2-13}
   & FPR & FNR & Acc & bAcc & F1 & Training time & FPR & FNR & Acc & bAcc & F1 & Training time \\ \midrule
  DNN          & 0.51 & 8.07 & 98.72 & 95.71 & 93.58 & 592   & 0.32  & 1.22 & 99.23 & 99.23 & 99.23 & 527   \\
  DNN$^+$      & 0.54 & 7.79 & 98.72 & 95.83 & 93.60 & 1559  & 0.05  & 0.83 & 99.54 & 99.56 & 99.56 & 1148  \\
  KDE          & 0.53 & 8.10 & 98.67 & 95.68 & 93.53 & 592   & 0.15  & 1.22 & 99.31 & 99.32 & 99.32 & 527   \\
  FD-VAE       & 1.14 & 23.3 & 96.62 & 87.79 & 82.12 & 1021  & 11.26 & 4.37 & 92.22 & 92.19 & 92.55 & 2182  \\
  AT-rFGSM$^k$ & 2.35 & 5.47 & 97.33 & 96.09 & 87.77 & 616   & 1.42  & 0.72 & 98.93 & 98.93 & 98.95 & 582   \\
  AT-Adam      & 4.01 & 5.47 & 95.84 & 95.26 & 82.14 & 1341  & 3.54  & 0.59 & 97.95 & 97.94 & 98.00 & 1018  \\
  PAD-SMA      & 1.70 & 5.94 & 97.87 & 96.18 & 89.93 & 21627 & 0.90  & 1.62 & 98.73 & 98.74 & 98.77 & 29837 \\
  MalPurifier  & 2.55 & 7.05 & 97.00 & 95.20 & 86.23 & 750   & 2.30  & 0.59 & 98.57 & 98.56 & 98.59 & 696   \\ \bottomrule
  \end{tabular}
\end{table*}

\noindent \textbf{Defenses considered for comparative analysis.} Note that all the defense methods consider DNN as the baseline classifier, and the SOTA defense mechanisms are as follows:
\begin{itemize}[leftmargin=*]
  \item \textbf{DNN}~\cite{grosse2017adversarial}. It employs a DNN model for malware detection without any countermeasures against evasion attacks.
  \item \textbf{DNN}$^+$~\cite{grosse2017statistical}. It enhances the robustness of the detector by another detector trained with an additional outlier class for detecting adversarial examples.
  \item \textbf{KDE}~\cite{pang2018towards}. It introduces a secondary detector that utilizes a \emph{Kernel Density Estimate} (KDE) method. This detector identifies adversarial examples in the final layer of the DNN that deviate significantly from normal data.
  \item \textbf{FD-VAE}~\cite{li2021robust}. It improves the DNN model by introducing an additional VAE for \emph{Feature Disentangle} (FD) in different classes and combining their detection outcomes to make the final decision (FD-VAE).  
  \item $\textbf{AT-rFGSM}^k$~\cite{al2018adversarial}. It strengthens the detector by \emph{Adversarial Training} (AT) with randomized rounding projection enabled $\emph{FGSM}^k$ attack (AT-rFGSM$^k$).
  \item \textbf{AT-Adam}~\cite{li2019enhancing}. It enhances the robustness of the DNN model via incorporating \emph{Adversarial Training} with the PGD attack optimized by Adam (AT-Adam).
  \item \textbf{PAD-SMA}~\cite{li2023pad}. It achieves \emph{Principled Adversarial Detection} by a DNN-based malware detector and an Input Convexity Neural Network (ICNN) based adversary detector, both of which are strengthened by adversarial training incorporating the \emph{Stepwise Mixture of Attacks} (PAD-SMA).
\end{itemize}

\noindent \textbf{Metrics.} The effectiveness of defenses is assessed using five standard metrics as follows. False Positive Rate (FPR) denotes the proportion of benign samples incorrectly classified as malicious and False Negative Rate (FNR) represents the proportion of malicious samples incorrectly classified as benign. Accuracy (Acc) is the percentage of the test examples that are correctly while balanced Accuracy (bAcc) can be defined as the average accuracy obtained on either class. F1 score denotes a harmonic mean of precision and recall that combines the performance of a classifier in terms of both false positives and false negatives. In addition, we include training time to evaluate the overhead of these methods.

\subsection{RQ1: Effectiveness and Cost without Attacks}\label{section5.1}
\noindent \textbf{Experimental Setup.} We compare MalPurifier with the aforementioned approaches on the two datasets. We use a DNN model with 2 fully-connected hidden layers (each having 200 neurons) with the ELU activation, and the other methods also use this architecture for malware detection. 

In detail, DNN$^+$~\cite{grosse2017statistical} leverages another detector hardened by adversarial training with the MaxMA attack against the DNN model to identify adversarial examples, and KDE~\cite{pang2018towards} relies on the close distance between activations to reject large manipulations without retraining. FD-VAE~\cite{li2021robust} incorporates a VAE-based indicator to classify clean data and adversarial examples, in which both the encoder and decoder consist of two layers (each layer having 600 neurons) with the Softplus activation. Moreover, AT-rFGSM$^k$~\cite{al2018adversarial} uses the PGD-$\ell_\infty$ attack, which has 50 iterations with step size 0.02, and AT-Adam~\cite{li2019enhancing} exploits the Adam optimizer with iterations 50, step size 0.02, and random starting point. PAD-SMA~\cite{li2023black} uses three attacks, including PGD-$\ell_1$ attack iterates 50 times, PGD-$\ell_2$ attack iterates 50 times with step size 0.5, and PGD-$\ell_\infty$ iterates 50 times with step size 0.02. 

The proposed method exploits a DAE-based purification model, which has two layers (each having 600 neurons) for both the encoder and decoder with the Sigmoid activation and introduces attention weights following the encoder. The training data of the DAE model is generated by Algorithm~\ref{algorithm1} with step size 0.01 and Algorithm~\ref{algorithm2} with noise level 0.001. We conduct a group of preliminary experiments and finally set $\alpha=\beta=0.5$ on both datasets. In addition, all detectors are tuned by the Adam optimizer with 100 epochs, batch size 128, and learning rate 0.001.

\noindent \textbf{Results.} Table~\ref{table1} exhibits the effectiveness and overhead of all detectors when there is no attack. We observe that DNN$^+$ achieves the highest detection accuracy (98.72\% on Drebin and 99.23\% on Androzoo) and F1 score (93.58\% on Drebin and 99.56\% on Androzoo), which are a little higher than those of the basic DNN model. The reason may be that adversarial training introduces extra adversarial examples that help DNN$^+$ identify more malicious samples, resulting in a lower FNR and higher F1 score.

An interesting observation is that KDE takes the same time as DNN whereas the other methods have higher overhead in the training phase. The reason may be that the KDE method builds another KDE-based detector without retraining, while the others train a separative model to detect adversarial examples. We further observe that MalPurifier's FNR decreases but FPR increases, leading to decreased accuracy (1.72\% on Drebin and 0.66\% on Androzoo) on clean data. which is similar to that of adversarial training methods (e.g., AT-rFGSM$^k$, AT-Adam, and PAD-SMA). 

Notably, the FPR of MalPurifier remains at a low and acceptable level (2.55\% on Drebin and 2.30\% on Androzoo), which is comparable to AT-Adam (4.01\% on Drebin) and lower than FD-VAE (11.26\% on Androzoo). This demonstrates that the protective noise injection strategy does not cause the purifier to misclassify benign apps. On the contrary, it teaches the purifier to distinguish the natural feature variations of benign samples from adversarial perturbations, thereby suppressing over-purification and keeping the FPR under control.

\begin{tcolorbox}[colback=gray!10,colframe=black,leftrule=0.5pt,rightrule=0.5pt,toprule=0.5pt,bottomrule=0.5pt,left=1pt,right=1pt,top=1pt,bottom=1pt]
  \textbf{Answer to RQ1:} MalPurifier exhibits a slight decrease in accuracy on clean data with a slightly increased overhead. In comparison, FD-VAE experiences a significant decrease in accuracy whereas PAD-SMA has an excessively long training time on both datasets.
\end{tcolorbox}

\begin{figure}[t]
  \centering
  \includegraphics[width=0.48\textwidth]{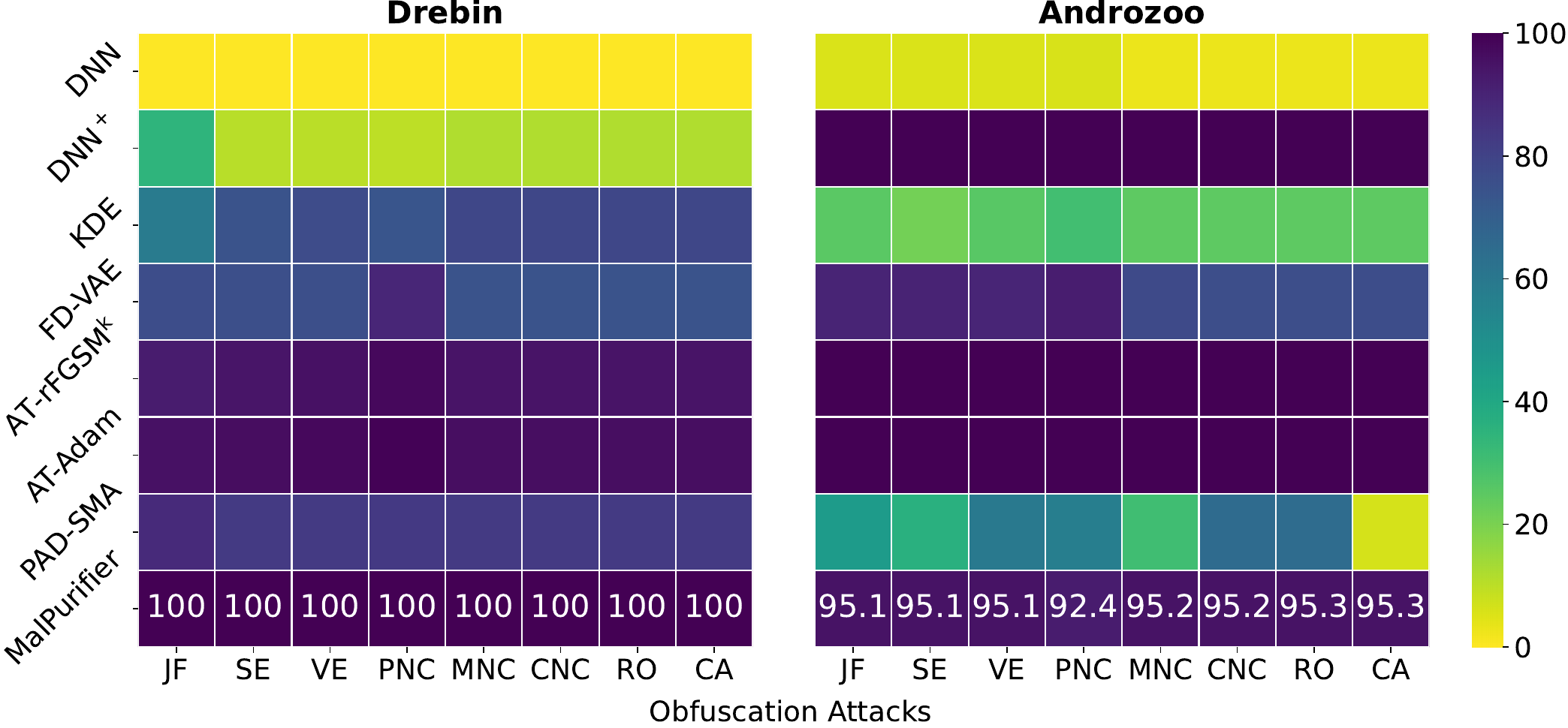} 
  \caption{The accuracy (\%) of different detectors against black-box attacks on Drebin and Androzoo datasets. The color gradient ranging from light to dark represents the increasing accuracy from low to high, with the effectiveness of MalPurifier against each attack annotated in the square.}
  \label{heatmap}
\end{figure}

\begin{figure}[t]
  \centering
  \includegraphics[width=0.48\textwidth]{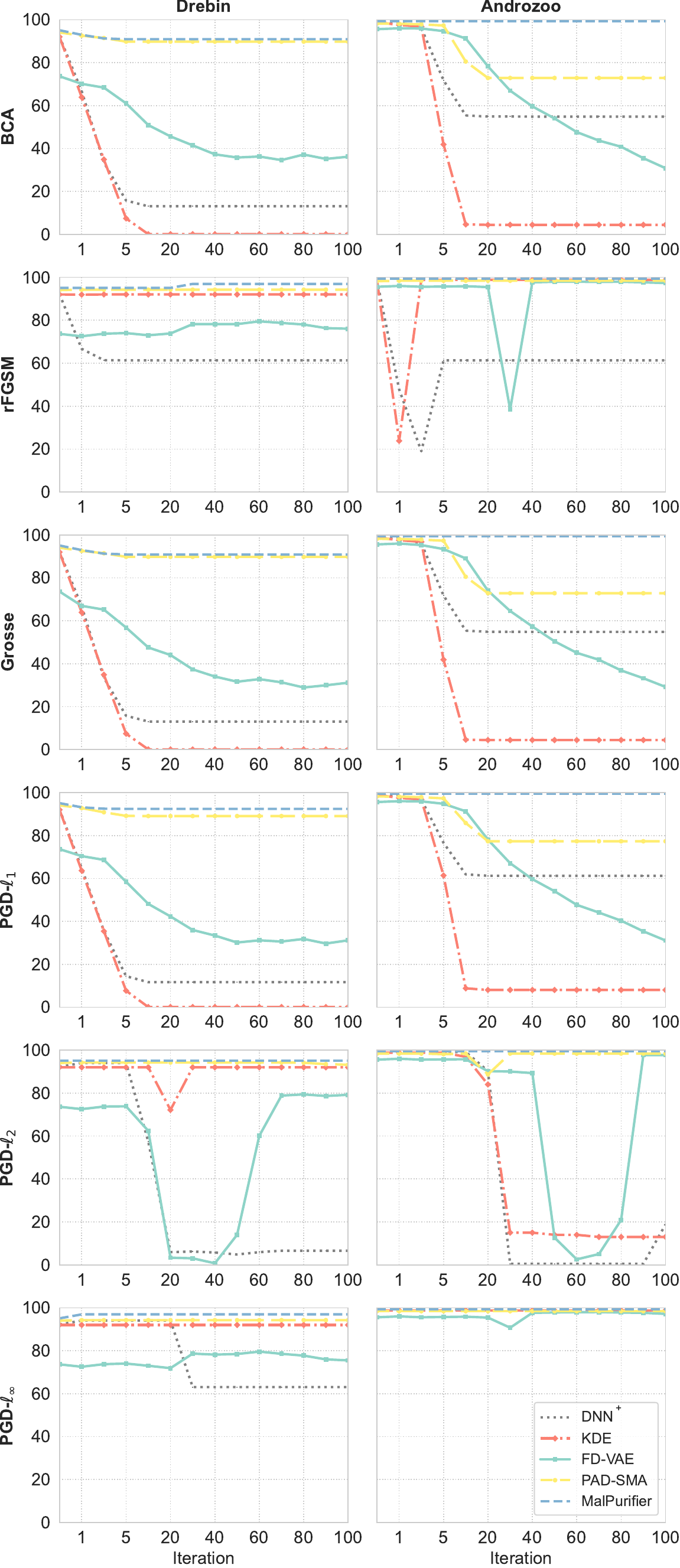} 
  \caption{The accuracy (\%) of different detectors against gradient-based grey-box attacks on Drebin (left panel) and Androzoo (right panel) datasets, along with the iteration ranging from 0 to 100.}
  \label{iteration}
\end{figure}

\subsection{RQ2: Robustness against Black-Box Attacks}\label{section5.2}
\noindent \textbf{Experimental Setup.} After establishing MalPurifier's baseline performance on clean data, we now investigate its effectiveness against black-box attacks. In detail, we measure the accuracy of all aforementioned methods under obfuscation attacks. These attacks utilize obfuscation technology to modify and conceal malicious functionality without the knowledge of the target classifier. Specifically, we utilize an obfuscator called AVPASS~\cite{jeon2017avpass}, to wage 8 kinds of attacks to perturb malware examples and extract features from the modified versions on the test set. Note that these obfuscation attacks are used exclusively for evaluation purposes and are never used to train MalPurifier. This design ensures that MalPurifier's robustness is not contingent on knowledge of the specific obfuscation technique employed by the attacker.

\begin{table*}[t]
  \caption{Accuracy (\%) of different defenses against gradient-free and ensemble-based grey-box attacks on Drebin and Androzoo datasets.}
  \centering
  \label{table2}
  \begin{tabular}{@{}l|ccccc|ccccc@{}}
  \toprule
  \multirow{2}{*}{Attack} & \multicolumn{5}{c|}{Drebin} & \multicolumn{5}{c}{Androzoo} \\ \cmidrule(l){2-11}
   & DNN$^+$ & KDE & FD-VAE & PAD-SMA & MalPurifier & DNN$^+$ & KDE & FD-VAE & PAD-SMA & MalPurifier \\ \midrule
  No Attack      & 92.21 & 91.93 & 73.19 & 94.06 & \cellcolor{gray!30}\textbf{95.08} & 98.74 & 98.83 & 95.59 & 98.20 & \cellcolor{gray!30}\textbf{99.41} \\
  Salt \& Pepper & 0.000 & 0.000 & 100.0 & 100.0 & \cellcolor{gray!30}\textbf{100.0} & 87.98 & 89.46 & 91.90 & \cellcolor{gray!30}\textbf{99.96} & 99.24 \\
  Pointwise      & 0.000 & 0.000 & 69.46 & 89.70 & \cellcolor{gray!30}\textbf{100.0} & 77.36 & 71.56 & 90.68 & 97.97 & \cellcolor{gray!30}\textbf{99.24} \\
  MaxMA          & 24.58 & 91.93 & 58.63 & 94.25 & \cellcolor{gray!30}\textbf{96.66} & 19.71 & 98.83 & 94.78 & 98.42 & \cellcolor{gray!30}\textbf{99.41} \\
  iMaxMA         & 24.58 & 91.93 & 58.63 & 94.25 & \cellcolor{gray!30}\textbf{96.66} & 19.71 & 98.83 & 94.78 & 98.42 & \cellcolor{gray!30}\textbf{99.41} \\
  StepwiseMA     & 12.71 & 0.649 & 13.82 & 89.05 & \cellcolor{gray!30}\textbf{96.66} & 61.21 & 8.236 & 18.23 & 77.32 & \cellcolor{gray!30}\textbf{99.41} \\
  AutoAttack     & 81.35 & 82.93 & 42.30 & 93.69 & \cellcolor{gray!30}\textbf{96.94} & 1.800 & 27.41 & 38.25 & 98.42 & \cellcolor{gray!30}\textbf{99.41} \\ \bottomrule
  \end{tabular}
\end{table*}

For Java Reflection (JF), this attack can hide public and static system APIs invoked in Smali using the reflection API. The encryption attacks typically encrypt the const-string and variable names in the decode, i.e., String Encryption (SE) and Variable Encryption (VE). The Package Name Change (PNC), Method Name Change (MNC), and Class Name Change (CNC) attacks change the names of packages, methods, and classes by replacing them with random characters, respectively. For the Resource Obfuscation (RO) attack, it changes pixel or adds one byte to the image files of APKs, along with the modification of related \texttt{AndroidManifest.xml}. Finally, we combine the above techniques to produce a Combined Attack (CA).

\noindent \textbf{Results.} Fig.~\ref{heatmap} illustrates the accuracy of the detectors on Drebin (left panel) and Androzoo (right panel) datasets under 8 obfuscation-based black-box attacks. We make the first observation that DNN can not defeat all these attacks (accuracy $\leqslant$ 0.344\% on Drebin and $\leqslant$ 5.871\% on Androzoo), demonstrating that such attacks can hide malicious features to evade detection. Nevertheless, attackers produce adversarial examples in a black-box manner, they cannot effectively evade these detectors except for the DNN model. 

We further observe significant differences between the robustness of some detectors against these attacks. For example, DNN$^+$ shows poor performance on Drebin whereas achieves high robustness on Androzoo. This can be attributed that the structures in the two datasets are different, and the data imbalance may lead to this as well. Another observation is that some adversarial training methods show high robustness on black-box attacks, which may be attributed to the similarity between adversarial examples generated by PGD attacks and obfuscation technology.

\begin{tcolorbox}[colback=gray!10,colframe=black,leftrule=0.5pt,rightrule=0.5pt,toprule=0.5pt,bottomrule=0.5pt,left=1pt,right=1pt,top=1pt,bottom=1pt]
  \textbf{Answer to RQ2:} MalPurifier outperforms the other methods against all black-box attacks on the Drebin dataset (accuracy of 100\%), and achieves accuracy $\geqslant$ 95\% against 7 black-box attacks on the Androzoo dataset.
\end{tcolorbox}

\subsection{RQ3: Robustness against Grey-Box Attacks}\label{section5.3}
\noindent \textbf{Experimental Setup.} We here evaluate Malpurifier's performance against grey-box attacks, where the adversaries are aware of the baseline DNN classifier but remain oblivious to other defensive mechanisms. Since DNN, AT-rFGSM$^k$, and AT-Adam lack additional detectors or indicators, we solely consider the robustness of DNN$^+$, KDE, FD-VAE, PAD-SMA, and MalPurifier against 12 grey-box attacks. 

First, we wage 6 gradient-based grey-box attacks in an oblivious manner on test malware examples. For BCA~\cite{al2018adversarial}, Grosse~\cite{hu2022generating}, and PGD-$\ell_1$~\cite{zhang2019interpreting} attacks, we perturb one feature per time with a maximum 100 iterations. For rFGSM~\cite{qiao2023adversarial} and PGD-$\ell_\infty$~\cite{zhang2019interpreting}, we iterate these attack algorithms with 100 iterations and a step size of 0.02. The PGD-$\ell_2$~\cite{zhang2019interpreting} attack is set with 100 iterations and a step size of 0.5.

We also incorporate 2 gradient-free attacks in the grey-box scenario. We conduct Salt \& Pepper attack~\cite{li2020adversarial} by increasing the noise intensity of 0.001 each time until misclassification and repeating this process 10 times. The Pointwise~\cite{vo2021query} attack utilizes Salt \& Pepper as the initial attack and minimizes the needed perturbations.

Furthermore, 4 ensemble-based grey-box attacks are included. We combine PGD-$\ell_1$, PGD-$\ell_2$, and PGD-$\ell_\infty$ to perform MaxMA~\cite{li2023pad} attack and run it 5 times with the random starting point for the iMaxMA~\cite{li2023pad} attack. We iterate the StepwiseMA~\cite{li2023pad} attack 100 times with a step size of 0.5 for PGD-$\ell_2$ and 0.02 for PGD-$\ell_\infty$, and the AutoAttack~\cite{croce2020reliable} comprises APGD-CE and FAB with the $\ell_2$ norm.

\noindent \textbf{Results.} Fig.~\ref{iteration} depicts the accuracy curves of these methods on Drebin (left panel) and Androzoo (right panel) datasets under 6 gradient-based grey-box attacks with the iteration from 0 to 100. We first observe an important observation that none of these attacks can evade MalPurifier (accuracy $\leqslant$ 90.91\% on Drebin and $\leqslant$ 99.41\% on Androzoo), demonstrating the high robustness of the proposed approach.

% \usepackage{booktabs}
% \usepackage{multirow}
\begin{table*}[htbp]
  \caption{Accuracy (\%) of different defenses under white-box attacks where adversaries know all defensive mechanisms (if applicable).}
  \centering
  \label{table3}
  \begin{tabular}{@{}l|cccccccc@{}}
  \toprule
  \multirow{2}{*}{Attack} & \multicolumn{8}{c}{Drebin} \\ \cmidrule(l){2-9}
   & DNN & AT-rFGSM$^k$ & AT-Adam & DNN$^+$ & KDE & FD-VAE & PAD-SMA & MalPurifier \\ \midrule
  BCA            & 0.000 & 6.122 & 42.21 & 0.000 & 0.000 & 5.102 & 80.15 & \cellcolor{gray!30}\textbf{90.91} \\
  rFGSM          & 0.000 & 14.94 & 85.81 & 59.46 & 91.93 & 74.40 & 94.25 & \cellcolor{gray!30}\textbf{95.08} \\
  Grosse         & 0.000 & 6.122 & 41.93 & 0.000 & 0.000 & 2.690 & 78.76 & \cellcolor{gray!30}\textbf{90.91} \\
  PGD-$\ell_1$   & 0.000 & 0.186 & 41.84 & 0.000 & 0.000 & 0.835 & 77.83 & \cellcolor{gray!30}\textbf{99.72} \\
  PGD-$\ell_2$   & 9.833 & 30.61 & 86.74 & 0.093 & 62.06 & 2.319 & 92.30 & \cellcolor{gray!30}\textbf{95.08} \\
  PGD-$\ell_\infty$ & 0.000 & 15.21 & 92.30 & 51.67 & 91.93 & 75.97 & 94.25 & \cellcolor{gray!30}\textbf{95.08} \\
  Salt \& Pepper & 0.000 & 97.96 & 99.07 & 0.000 & 0.000 & 80.43 & 95.55 & \cellcolor{gray!30}\textbf{100.0} \\
  Pointwise      & 0.000 & 94.43 & 94.53 & 0.000 & 0.000 & 67.97 & 87.65 & \cellcolor{gray!30}\textbf{99.72} \\
  MaxMA          & 0.000 & 0.371 & 45.83 & 0.000 & 0.000 & 1.299 & 77.37 & \cellcolor{gray!30}\textbf{95.08} \\
  iMaxMA         & 0.000 & 0.371 & 45.83 & 0.000 & 0.000 & 1.299 & 77.83 & \cellcolor{gray!30}\textbf{95.08} \\
  StepwiseMA     & 0.000 & 0.371 & 45.55 & 0.371 & 25.05 & 1.206 & 88.22 & \cellcolor{gray!30}\textbf{95.08} \\
  AutoAttack     & 0.000 & 78.39 & 71.71 & 81.35 & 82.93 & 42.30 & 93.69 & \cellcolor{gray!30}\textbf{96.94} \\
  Orth PGD-$\ell_1$      & $-$ & $-$ & $-$ & 0.000 & 6.401 & 12.43 & 94.25 & \cellcolor{gray!30}\textbf{95.08} \\
  Orth PGD-$\ell_2$      & $-$ & $-$ & $-$ & 3.711 & 0.000 & 17.90 & 94.25 & \cellcolor{gray!30}\textbf{95.08} \\
  Orth PGD-$\ell_\infty$ & $-$ & $-$ & $-$ & 46.01 & 89.98 & 55.29 & 94.25 & \cellcolor{gray!30}\textbf{95.08} \\
  Orth MaxMA             & $-$ & $-$ & $-$ & 0.000 & 0.000 & 13.08 & 94.25 & \cellcolor{gray!30}\textbf{95.08} \\
  Orth iMaxMA            & $-$ & $-$ & $-$ & 0.000 & 0.000 & 13.08 & 94.25 & \cellcolor{gray!30}\textbf{95.08} \\   \midrule
  \multirow{2}{*}{Attack} & \multicolumn{8}{c}{Androzoo} \\ \cmidrule(l){2-9}
   & DNN & AT-rFGSM$^k$ & AT-Adam & DNN$^+$ & KDE & FD-VAE & PAD-SMA & MalPurifier \\ \midrule
  BCA            & 0.000 & 0.090 & 35.01 & 0.000 & 10.62 & 30.56 & 86.72 & \cellcolor{gray!30}\textbf{99.37} \\
  rFGSM          & 0.000 & 8.776 & 98.16 & 6.841 & 98.83 & 96.00 & 98.43 & \cellcolor{gray!30}\textbf{99.41} \\
  Grosse         & 0.000 & 0.090 & 35.01 & 0.000 & 0.045 & 28.04 & 56.30 & \cellcolor{gray!30}\textbf{99.37} \\
  PGD-$\ell_1$   & 0.000 & 0.000 & 15.26 & 0.000 & 0.585 & 3.465 & 40.41 & \cellcolor{gray!30}\textbf{99.41} \\
  PGD-$\ell_2$   & 62.96 & 89.96 & 93.70 & 89.83 & 73.58 & 90.14 & \cellcolor{gray!30}\textbf{98.34} & 97.75 \\
  PGD-$\ell_\infty$ & 0.000 & 90.19 & 93.34 & 18.59 & 98.83 & 97.71 & 98.42 & \cellcolor{gray!30}\textbf{99.41} \\
  Salt \& Pepper & 89.47 & 99.73 & \cellcolor{gray!30}\textbf{99.82} & 87.17 & 89.06 & 91.85 & 98.38 & 99.41 \\
  Pointwise      & 71.56 & 99.25 & 99.33 & 76.50 & 68.36 & 90.68 & 94.27 & \cellcolor{gray!30}\textbf{99.37} \\
  MaxMA          & 0.000 & 0.000 & 14.94 & 0.000 & 0.585 & 7.111 & 40.41 & \cellcolor{gray!30}\textbf{97.44} \\
  iMaxMA         & 0.000 & 0.000 & 14.94 & 0.000 & 0.585 & 6.391 & 40.41 & \cellcolor{gray!30}\textbf{97.44} \\
  StepwiseMA     & 0.000 & 0.000 & 16.11 & 0.000 & 79.16 & 3.555 & 76.73 & \cellcolor{gray!30}\textbf{99.41} \\
  AutoAttack     & 0.540 & 98.25 & 97.80 & 1.800 & 27.41 & 38.25 & 98.42 & \cellcolor{gray!30}\textbf{99.41} \\
  Orth PGD-$\ell_1$      & $-$ & $-$ & $-$ & 0.000 & 0.000 & 8.731 & 98.42 & \cellcolor{gray!30}\textbf{99.41} \\
  Orth PGD-$\ell_2$      & $-$ & $-$ & $-$ & 0.045 & 78.22 & 43.70 & 98.42 & \cellcolor{gray!30}\textbf{99.41} \\
  Orth PGD-$\ell_\infty$ & $-$ & $-$ & $-$ & 19.13 & 93.65 & 97.21 & 98.42 & \cellcolor{gray!30}\textbf{99.41} \\
  Orth MaxMA             & $-$ & $-$ & $-$ & 0.000 & 0.000 & 7.111 & 98.42 & \cellcolor{gray!30}\textbf{99.41} \\
  Orth iMaxMA            & $-$ & $-$ & $-$ & 0.000 & 0.000 & 7.111 & 98.42 & \cellcolor{gray!30}\textbf{99.41} \\ \bottomrule
  \end{tabular}
\end{table*}

There is a decreasing trend on the curves of DNN$^+$, KDE, and PAD-SMA against BCA, Grosse, and PGD-$\ell_1$ attacks until 20 iterations. This is because such attacks in an oblivious manner will stop manipulations when the adversarial example can evade detection. Moreover, there exists a dip in accuracies of DNN$^+$, KDE, and FD-VAE against rFGSM and PGD-$\ell_2$ attacks. An interesting observation is that KDE can effectively mitigate PGD-$\ell_\infty$ attack whereas fails to defeat other attacks. The reason may be that KDE relies on the close distance between activations to reject large manipulations that are used by the PGD-$\ell_\infty$ attack, while the basic DNN model is sensitive to small perturbations, leading to the failure against other attacks. Another observation is that the robust accuracies of FD-VAE against all these gradient-based attacks are very different on the two datasets. The reason may be that the threshold of reconstruction error in FD-VAE relies on the distribution of training datasets, which may be significantly different from each other.

Table~\ref{table2} reports the results of Salt $\&$ Pepper, Pointwise, MaxMA, iMaxMA, StepwiseMA, and Autoattack, which are not suitable for number of iterations. We first observe that MalPurifier can achieves the highest accuracy against them, except for the Salt $\&$ Pepper attack with 99.24\% accuracy on Androzoo. We further observe that there exist significant differences in the accuracy of all defensive methods when dealing with various attacks, except for MalPurifier. For example, KDE can mitigate MaxMA and iMaxMA with an accuracy of 91.93\% but cannot defeat StepwiseMA attack with an accuracy of 0.649\% on Drebin. This is because other methods are vulnerable to large or small manipulations, while the diversified perturbation mechanism can help Malpurifier defend against a range of perturbations.

\begin{tcolorbox}[colback=gray!10,colframe=black,leftrule=0.5pt,rightrule=0.5pt,toprule=0.5pt,bottomrule=0.5pt,left=1pt,right=1pt,top=1pt,bottom=1pt]
  \textbf{Answer to RQ3:} MalPurifier is significantly more robust than DNN$^+$, KDE, FD-VAE, and PAD-SMA. It achieves an accuracy of $\geqslant$ 90.91\% and $\geqslant$ 99.24\% on both datasets against grey-box attacks, wherein all of these adversarial examples are previously unseen by the model.
\end{tcolorbox}

\subsection{RQ4: Robustness against White-Box Attacks}\label{section5.4}
\noindent \textbf{Experimental Setup.} To further evaluate MalPurifier under worst-case conditions, we now progress to white-box attacks, where attackers have complete knowledge of both the detector $f$ and the adversarial indicator or purifier $g$.

First, we adapt the 12 grey-box attacks to white-box attacks by solving the problem in Eq.~\ref{eq7}. Since DNN, AT-rFGSM$^k$, and AT-Adam do not have any adversarial indicator, the grey-box attacks aimed at these defenses can trivially fulfill the adaptive requirement of white-box attacks.

Furthermore, we improve the other 5 white-box attacks by producing perturbations into two components (e.g., detector $f$ and purifier $g$) in an "orthogonal" (dubbed Orth) manner~\cite{bryniarski2022evading} to prevent perturbation waste, including Orth PGD-$\ell_1$, PGD-$\ell_2$, PGD-$\ell_\infty$, MaxMA, and iMaxMA. For similar reasons, these attacks are not applicable to DNN, AT-rFGSM$^k$, and AT-Adam approaches. Note that we utilize the same hyper-parameters as those in Section~\ref{section5.3}, except for PGD-$\ell_1$ with 500 iterations, PGD-$\ell_2$ with 200 iterations and step size 0.05, and PGD-$\ell_\infty$ with 500 iterations and step size 0.002 in all PGD-based attacks.

\noindent \textbf{Results.} As shown in Table~\ref{table3}, we can first observe that DNN is very vulnerable to all attacks, with 0\% accuracy against 11 attacks on the Drebin dataset and 8 attacks on the Androzoo dataset. However, the Salt $\&$ Pepper and Pointwise attacks achieve the lowest attack effectiveness in evading DNN on Androzoo, because both of them modify malware examples without using the internal information of target detectors.

Second, adversarial training methods (e.g., AT-rFGSM$^k$, and AT-Adam) can harden the robustness of DNN to some extent. For example, AT-rFGSM$^k$ can mitigate the Salt $\&$ Pepper, Pointwise, and AutoAttack attacks, and AT-Adam is also effective in defeating rFGSM, PGD-$\ell_2$ and PGD-$\ell_\infty$ attacks. Nevertheless, they are still sensitive to BCA, Grosse, PGD-$\ell_1$, MaxMA, iMaxMA, and StepwiseMA attacks (with an accuracy $\leqslant$ 45.83\% on both datasets) that are unseen previously. These results indicate the limitations of adversarial training methods in terms of generalization.

Third, although DNN$^+$, KDE, and FD-VAE incorporate another adversary detector, they show limited effectiveness under a few attacks, and suffer from unseen attacks such as PGD-$\ell_1$, MaxMA, iMaxMA, Orth PGD-$\ell_1$, Orth MaxMA, and Orth iMaxMA (with an accuracy $\leqslant$ 13.08\% on both datasets). Especially, PAD-SMA achieves the highest accuracy of 98.34\% against the PGD-$\ell_2$ attack on Androzoo. However, PAD-SMA is still sensitive to the Grosse, PGD-$\ell_1$, MaxMA, and iMaxMA attacks (with an accuracy $\leqslant$ 78.76\% on Drebin and $\leqslant$ 56.30\% on Androzoo). Considering its high time overhead reported in Table~\ref{table1}, it cannot be called a perfect solution.

In summary, MalPurifier significantly outperforms other defenses, achieving the highest accuracy against all 17 white-box attacks on the Drebin dataset and 15 attacks on the Androzoo dataset. This indicates that the purification model can accurately recover the original forms of adversarial examples even if the adversary knows its existence.

\begin{tcolorbox}[colback=gray!10,colframe=black,leftrule=0.5pt,rightrule=0.5pt,toprule=0.5pt,bottomrule=0.5pt,left=1pt,right=1pt,top=1pt,bottom=1pt]
  \textbf{Answer to RQ4:} MalPurifier outperforms other defenses in the condition that adversaries know all defensive mechanisms, and significantly hardens the malware detector against a wide range of white-box attacks (with an accuracy $\geqslant$ 90.91\% on Drebin and $\geqslant$ 97.44\% on Androzoo).
\end{tcolorbox}

\begin{figure}[t]
  \centering
  \includegraphics[width=0.48\textwidth]{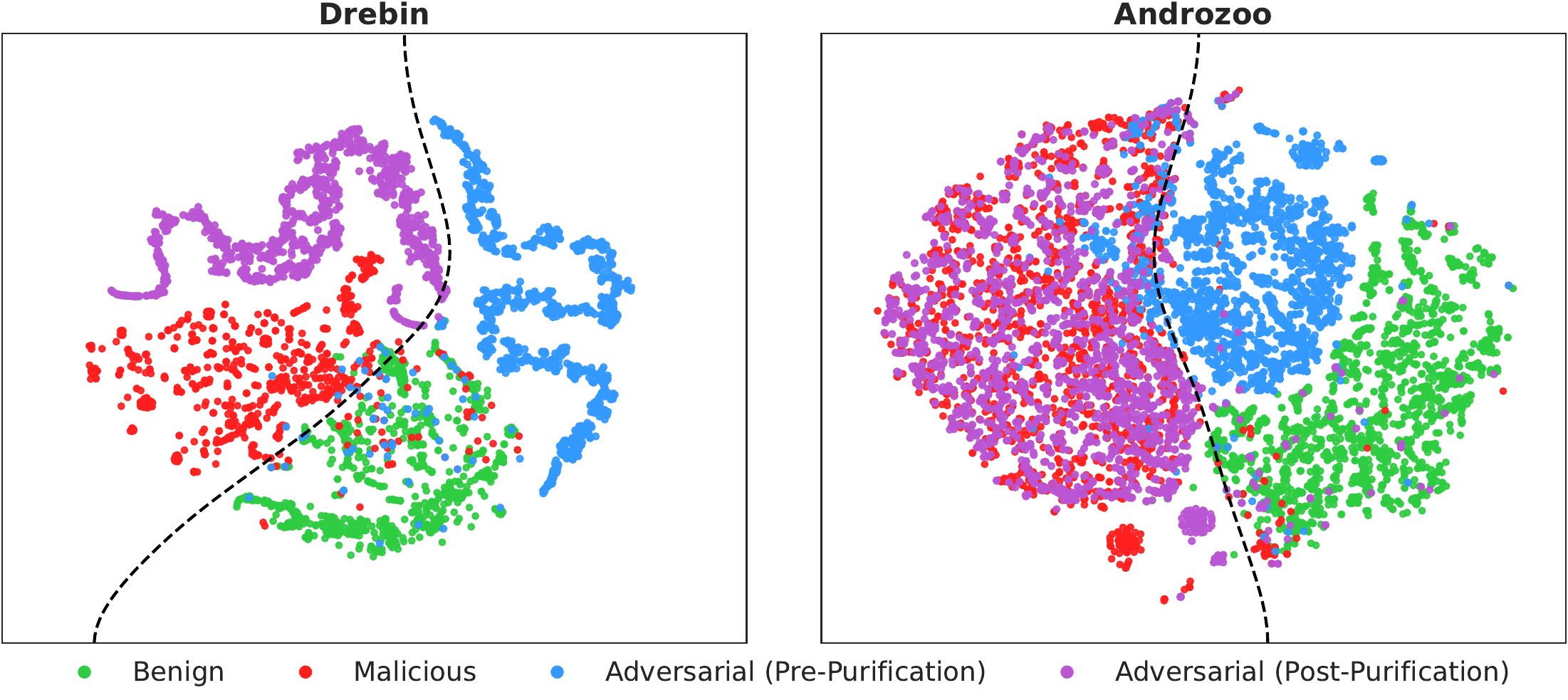} 
  \caption{t-SNE visualization of benign, clean malware, adversarial, and purified feature representations under the PGD-$\ell_2$ white-box attack on the Drebin (left) and Androzoo (right) datasets.}
  \label{tsne}
\end{figure}

\begin{figure}[t]
  \centering
  \includegraphics[width=0.48\textwidth]{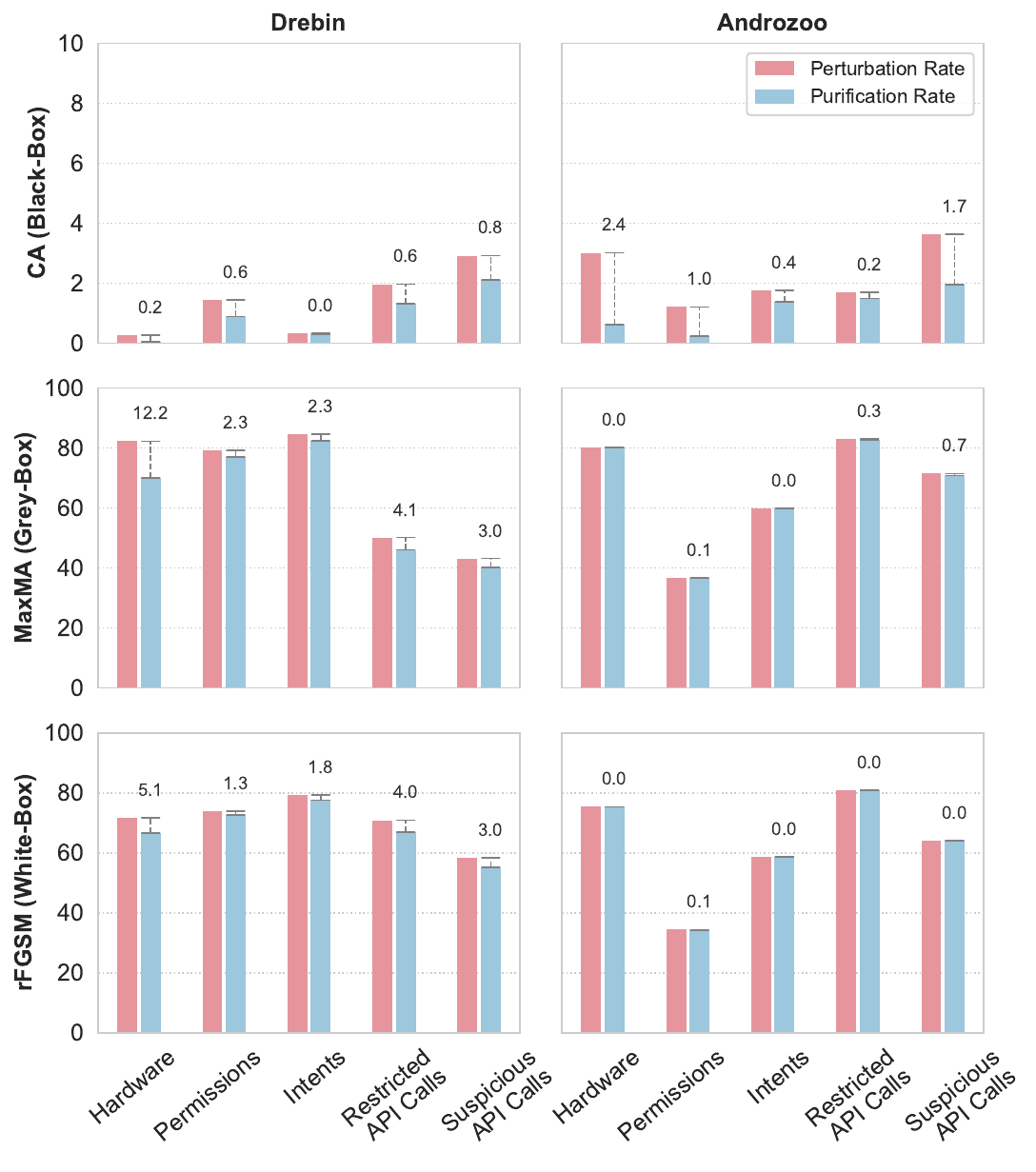} 
  \caption{Feature-level analysis of perturbation and purification rates (\%) per feature category under representative attacks.}
  \label{feature_analysis}
\end{figure}

\subsection{RQ5: Interpretability of the Purification Process}
\noindent \textbf{Experimental Setup.} To provide interpretability analyses to understand how the purification process corrects adversarial perturbations and which features are most relevant to this correction, we here conduct a global-level visualization and a feature-level analysis.

For the global-level analysis, we select the PGD-$\ell_2$ white-box attack as a representative adversarial threat, using the same parameters as in Section~\ref{section5.4}. On each dataset, we apply t-SNE to project benign samples, malware samples, and adversarial samples with their purified versions into a two-dimensional space for visualization.

For the feature-level analysis, we select three attack scenarios: a black-box CA attack (configured as in Section~\ref{section5.2}), a grey-box MaxMA attack (configured as in Section~\ref{section5.3}), and a white-box rFGSM attack (configured as in Section~\ref{section5.4}). For each attack on each dataset, we compute the perturbation rate and purification rate for each feature category across all test malware samples.

\noindent \textbf{Results.} Fig.~\ref{tsne} presents the t-SNE visualization of feature representations on both datasets. In both panels, the benign samples and malware samples form clearly separated clusters, confirming that the original feature space provides discriminative information for classification.

After the PGD-$\ell_2$ attack, the adversarial samples are significantly displaced from the malware cluster, with many points drifting toward or even overlapping with the benign region, causing the classifier to misidentify them. Critically, the purified samples are pulled back on both datasets, where the purified points concentrate near the malware region. These results provide intuitive evidence that MalPurifier's purification process successfully restores adversarial samples to the malicious data manifold, thereby enabling the downstream classifier to correctly recognize them.

Fig.~\ref{feature_analysis} presents the per-category perturbation and purification rates under three representative attacks. Under the black-box CA attack, perturbation rates remain below 4\% across all categories, and MalPurifier restores nearly all perturbed features with residual gaps $\leqslant$ 2.4\%. Under the stronger MaxMA and rFGSM attacks on Drebin, perturbation rates rise sharply to at most 85\%, yet the purifier still closely tracks them. The largest residual gap is 12.2\% for Hardware features under MaxMA, while Permissions, Intents, and the two API categories all stay within 4.1\%. On Androzoo, purification is even more thorough, indicating near-perfect feature restoration. Overall, MalPurifier consistently corrects the vast majority of adversarial perturbations across diverse feature categories and attack scenarios.

\begin{tcolorbox}[colback=gray!10,colframe=black,leftrule=0.5pt,rightrule=0.5pt,toprule=0.5pt,bottomrule=0.5pt,left=1pt,right=1pt,top=1pt,bottom=1pt]
  \textbf{Answer to RQ5:} The t-SNE visualization confirms that MalPurifier effectively restores adversarial samples at the representation level. The per-feature analysis further shows that MalPurifier corrects perturbations across all feature categories with residual gaps of at most 12.2\% on Drebin and 2.4\% on Androzoo, achieving near-perfect restoration under white-box attacks on Androzoo.
\end{tcolorbox}

\subsection{RQ6: Advantage and Transferability of the Purifier}
\noindent \textbf{Experimental Setup.} While previous experiments have demonstrated MalPurifier's effectiveness against increasingly sophisticated attacks, we now investigate two critical aspects: (i) whether our choice of DAE as the purification model is optimal compared to alternative techniques, and (ii) whether MalPurifier can be flexibly integrated with different detection architectures.

First, to justify the effectiveness of the DAE-based purification model, we compare it with alternative purification techniques, including VAE, PCA, GAN, and DM. The DAE model is trained with the same hyper-parameters as that in Section~\ref{section5.1}. VAE shares the same architecture of the encoder and decoder as DAE, but with the Softplus activation, and it introduces a Kullback–Leibler (KL) weight of 1.0 and regularization coefficient of 0.001. For PCA, the number of principal components is selected to retain more than 95\% of the variance, and the projection matrix is derived from the training data. Both generator and discriminator of the GAN model use a hidden size of 600, and the latent dimension is set to 256 with a gradient penalty of 10. DM uses a sinusoidal time embedding with a dimension of 256, and a two-layer \emph{Multilayer Perceptron} (MLP) with a hidden size of 600 to predict noise residuals at arbitrary steps.

For comprehensive comparison among different purification methods, we select several test scenarios including a clean setting and five white-box attacks: rFGSM, PGD-$\ell_\infty$, Salt \& Pepper, MaxMA, and StepwiseMA. Attack configurations follow the setup described in Section~\ref{section5.4}, and we use a DNN-based model as the classifier with the same hyper-parameters as that in Section~\ref{section5.1}.

Second, to further evaluate the flexibility and transferability of the purification model, we also package the DAE model as a plug to protect other detectors, such as \emph{Support Vector Machine} (SVM), \emph{Fully Convolutional Network} (FCN), \emph{Long Short Term Memory} (LSTM), and \emph{Recurrent Neural Network} (RNN). In detail, We use an SVM model with the Sigmoid activation and an FCN model with 3 fully-connected hidden layers (each having 512, 256, and 128 neurons) with the ReLU activation. The LSTM model has a hidden layer with 200 neurons, a sequence length of 1, and uses the Sigmoid function as the activation. Furthermore, we build the RNN model with 3 hidden layers (each having 200 neurons) and the Sigmoid activation. All these models are tuned by the Adam optimizer with 100 epochs, batch size 128, and learning rate 0.001.

Note that the DAE-based purification model works as a plug-and-play preprocessing method, in which we do not retrain the model but directly apply it from the aforementioned experiments. In addition, we wage 6 attacks on test malware examples in a white-box manner, including BCA, PGD-$\ell_1$, PGD-$\ell_2$, PGD-$\ell_\infty$, Pointwise, and StepwiseMA, with the same hyper-parameters in Section~\ref{section5.4}.

\begin{figure*}[t]
  \centering
  \includegraphics[width=\textwidth]{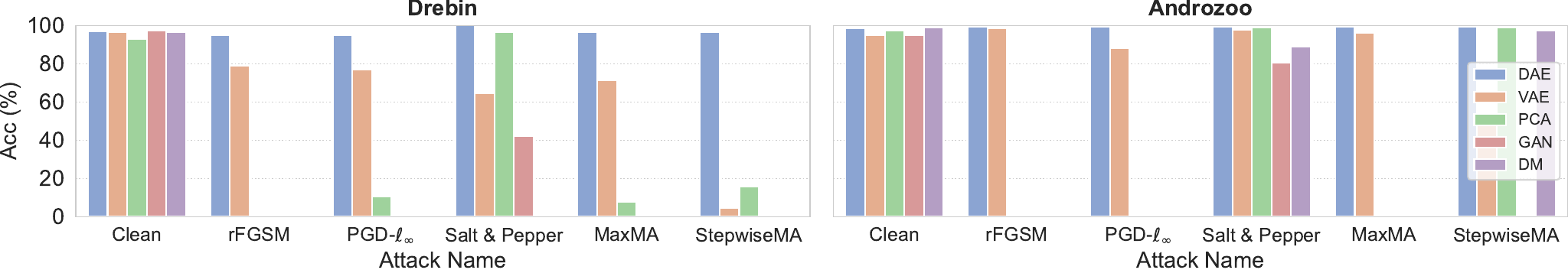} 
  \caption{The accuracy of different purification algorithms when equipped with the DNN-based classifier in the absence and presence of evasion attacks on Drebin and Androzoo datasets.}
  \label{purifier}
\end{figure*}

\begin{figure*}[t]
  \centering
  \includegraphics[width=\textwidth]{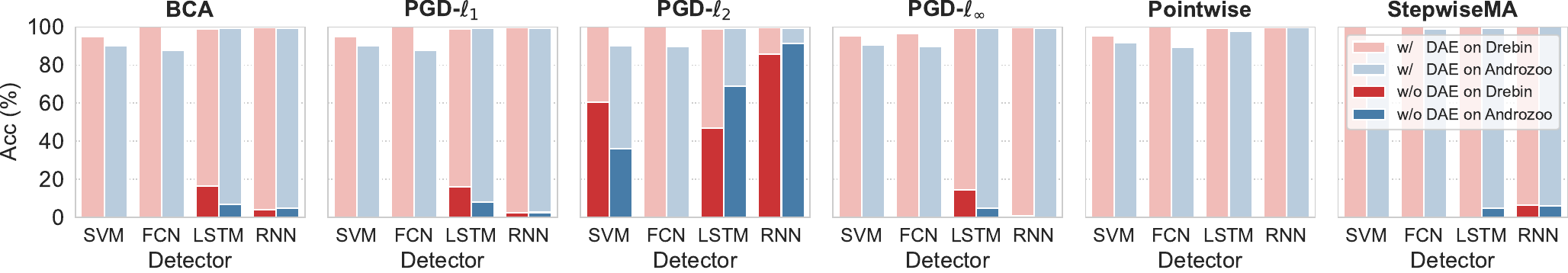} 
  \caption{The accuracy of different detectors against various evasion attacks when equipped with (w/) or without (w/o) the DAE-based purification model on Drebin (drawn in red) and Androzoo (drawn in blue) datasets.}
  \label{plug}
\end{figure*}

\noindent \textbf{Results.} Fig.~\ref{purifier} presents the classification accuracy of different purification methods on the Drebin and Androzoo datasets under both clean and adversarial conditions. As we can see, all methods achieve high accuracy on clean data, with minor differences among them. DAE achieves the accuracy of 97.00\% and 98.57\% on Drebin and Androzoo, respectively, which is slightly lower than that of GAN (97.27\%) on Drebin and DM (98.89\%) on Androzoo. Although DAE does not always achieve the highest accuracy on clean data, it consistently delivers stable and competitive results.

However, under adversarial attacks, DAE exhibits a clear and substantial advantage over alternative approaches. On both Drebin and Androzoo datasets, the proposed purification model consistently maintains high accuracy across all attack types, with only minimal performance degradation compared to the clean setting. In detail, On Drebin, DAE maintains accuracy $\geqslant$ 95.08\% across all attack types and achieves perfect accuracy of 100\% under the Salt \& Pepper attack, whereas other methods such as PCA, GAN, and DM experience dramatic drops in performance, with accuracy falling to even 0\% under several attacks. DAE remains highly robust on Androzoo, achieving accuracy $\geqslant$ 99.24\% for all adversarial attacks, while the competing methods show significant degradation, particularly under rFGSM, PGD-$\ell_\infty$, and MaxMA attacks. 

These results demonstrate that DAE not only provides stable and competitive performance on clean data but also offers superior robustness and generalizability against a wide range of adversarial perturbations, making it a highly effective purification model for defending Android malware detectors against diverse evasion attacks.

Moreover, Fig.~\ref{plug} depicts the accuracy improvement of other classifiers equipped with the DAE model on Drebin (drawn in red) and Androzoo (drawn in blue) under 6 white-box attacks. We make three observations as follows.

First, all detectors without enhancement cannot mitigate these attacks (with accuracy $\leqslant$ 16.42\% on Drebin and $\leqslant$ 8.19\% on Androzoo), except for the PGD-$\ell_2$ attack. This is because ML-based detectors are very vulnerable to these evasion attacks with small perturbations. Especially, the attack effectiveness of the Pointwise attack is 100\% when adversaries wage attacks on original detectors. Nevertheless, the PGD-$\ell_2$ attack, if running with a lot of iterations, will produce larger perturbations that may not evade detection.

Second, the DAE-based purification model works very well and can significantly improve the robustness of these detectors as a security plug (accuracy increase by $\geqslant$ 39.61\% for SVM, $\geqslant$ 87.71\% for FCN, $\geqslant$ 30.47\% for LSTM, and $\geqslant$ 7.97\% for RNN). Significantly, it boosts the accuracy of the RNN model against the Pointwise attack from 0\% to 99.72\% on Drebin and from 0\% to 99.82\% on Androzoo, rendering the previously vulnerable RNN model robust against this specific attack. These results strongly demonstrate the versatility of the DAE model, as it can seamlessly transfer to other models without the need for retraining.

Third, the accuracy values of these detectors are similar against different evasion attacks when equipped with the DAE model. The underlying reason for the observation is that the DAE model is trained in an independent and unsupervised way, and can accurately return the adversarial examples to their original forms. Hence, the effectiveness of equipping this security plug relies more on the detector itself, as the DAE model solely preprocesses the input data while the detector is responsible for classification.

\begin{tcolorbox}[colback=gray!10,colframe=black,leftrule=0.5pt,rightrule=0.5pt,toprule=0.5pt,bottomrule=0.5pt,left=1pt,right=1pt,top=1pt,bottom=1pt]
  \textbf{Answer to RQ6:} Compared with alternative purification techniques, the DAE model achieves the best trade-off between effectiveness, robustness, and generalizability under various adversarial attacks. Furthermore, it can be flexibly transferred as a plug-and-play module to enhance the robustness of different detectors, significantly improving their resistance to evasion attacks without retraining.
\end{tcolorbox}

\subsection{RQ7: Generalizability to Structural Attacks}\label{section5.7}
\noindent \textbf{Experimental Setup.} In real-world environments, adversaries may employ more sophisticated attacks that target structural or semantic properties~\cite{zapzalka2025semantics,abusnaina2022dl} of Android applications. To answer this research question, we evaluate MalPurifier's performance against these advanced attacks, thus providing a more rigorous test of its generalizability.

To evaluate MalPurifier's generalizability beyond the Drebin feature set used in RQ1–RQ6, we adopt a graph-based feature representation in this experiment following the MaMaDroid~\cite{onwuzurike2019mamadroid} methodology. The base model for all defenses continues to be a DNN, with its architecture and hyper-parameters remaining consistent with previous experiments. For MalPurifier, the DAE architecture remains unchanged, and we augmented its DAE training set with a small set (500 samples) of adversarial examples generated via the obfuscation-based attacks detailed in RQ2. Additionally, the step size used for generating diversified adversarial perturbations (Algorithm~\ref{algorithm1}) was set to 0.0001.

To ensure a fair comparison, all attack methods draw from the same set of possible manipulations sets (e.g., component injection, permission modification). Crucially, if a perturbation leads to an application crash or packaging failure during the attack process, the attempt is considered an attack failure, ensuring that only functional adversarial examples are evaluated. We then assess the defenses against four advanced structural attacks as follows:
\begin{itemize}[leftmargin=*]
    \item \textbf{Random Attack (RA)}: A black-box attack that iteratively and uniformly samples perturbations from the malware perturbation set to inject, and queries the target model with the perturbed sample. Our implementation follows the instructions in Ref.~\cite{he2023efficient}.
    \item \textbf{MAB}~\cite{song2022mab}: A reinforcement learning-based black-box attack, originally for Windows PE malware and adapted here for Android. It formulates the attack as a multi-armed bandit (MAB) problem to balance exploitation and exploration of manipulations. We implemented it following its official code\footnote{\url{https://github.com/weisong-ucr/MAB-malware}}.
    \item \textbf{AdvDroidZero}~\cite{he2023efficient}: A black-box query-based evasion attack framework designed for a zero-knowledge setting. It requires no prior information about the target model's features, architecture, or parameters. We implemented it with its official source code\footnote{\url{https://github.com/gnipping/AdvDroidZero-Access-Instructions}}.
    \item \textbf{EvadeDroid}~\cite{bostani2024evadedroid}: A problem-space black-box attack that iteratively injects transformations into malware. It leverages an n-gram based approach to find opcode-level similarities with benign applications, and we implemented it with its official code\footnote{\url{https://github.com/HamidBostani2021/EvadeDroid}}.
    \item \textbf{HRAT}~\cite{zhao2021structural}: A white-box structural attack specifically targeting graph-based malware detectors. It integrates a heuristic optimization model with reinforcement learning, performing four types of graph modifications that preserve functionality in the app's bytecode. We used its official implementation\footnote{\url{https://github.com/zacharykzhao/HRAT}} in our experiments.
\end{itemize}
Note that the query limit was set to 10 for query-based attacks (RA, MAB, AdvDroidZero, EvadeDroid). For HRAT, which focuses on modification counts, the maximum number of allowed modifications was 50.

\begin{table*}[t]
\centering
\caption{Accuracy (\%) of different defenses against advanced structural attacks on Drebin and Androzoo datasets.}
\label{table4}
\begin{tabular}{@{}l|ccccc|ccccc@{}}
\toprule
\multirow{2}{*}{Defense} & \multicolumn{5}{c|}{Drebin} & \multicolumn{5}{c}{Androzoo} \\ \cmidrule(l){2-11}
 & RA & MAB & AdvDroidZero & EvadeDroid & HRAT 
 & RA & MAB & AdvDroidZero & EvadeDroid & HRAT \\ \midrule
DNN          & 8.075 & 15.71 & 10.40 & 14.38 & 15.36 & 7.848 & 15.55 & 16.42 & 15.08 & 10.14 \\
AT-rFGSM$^k$ & 15.40 & 13.81 & 11.38 & 19.08 & 16.39 & 14.08 & 17.08 & 18.08 & 17.07 & 12.38 \\
AT-Adam      & 9.880 & 7.708 & 15.94 & 21.31 & 15.36 & 15.91 & 17.61 & 19.04 & 19.38 & 10.14 \\
DNN$^+$      & 27.31 & 40.84 & 34.08 & 40.95 & 56.44 & 29.08 & 37.92 & 33.38 & 30.08 & 49.64 \\
KDE          & 10.19 & 21.07 & 12.38 & 15.04 & 20.05 & 15.81 & 15.71 & 19.30 & 17.90 & 15.37 \\
FD-VAE       & 64.70 & 70.07 & 60.44 & 73.74 & 61.74 & 70.09 & 74.41 & 72.82 & 83.51 & 45.52 \\
PAD-SMA      & 47.07 & 56.71 & 50.55 & 60.90 & 70.37 & 56.38 & 64.05 & 60.60 & 75.37 & 63.55 \\
MalPurifier  & \cellcolor{gray!30}\textbf{79.74} & \cellcolor{gray!30}\textbf{90.88} & \cellcolor{gray!30}\textbf{87.85} & \cellcolor{gray!30}\textbf{92.38} & \cellcolor{gray!30}\textbf{92.08} 
             & \cellcolor{gray!30}\textbf{78.71} & \cellcolor{gray!30}\textbf{92.04} & \cellcolor{gray!30}\textbf{81.31} & \cellcolor{gray!30}\textbf{94.38} & \cellcolor{gray!30}\textbf{81.31} \\ \bottomrule
\end{tabular}
\end{table*}

\noindent \textbf{Results.} Table~\ref{table4} presents a detailed comparison of defense mechanisms against advanced structural attacks on both Drebin and Androzoo datasets. The results reveal several important trends as follows.

First, most defenses, except MalPurifier, struggle to provide adequate protection against these sophisticated attacks. For example, adversarially trained models such as AT-rFGSM$^k$ and AT-Adam achieve only 7.71\% and 13.81\% accuracy under MAB attacks on Drebin dataset, respectively. These models also remains similarly low across other attacks and datasets.

Second, auxiliary defenses such as DNN$^+$ and KDE offer only marginal improvements. For instance, DNN$^+$ achieves up to 56.44\% accuracy against HRAT on Drebin, but its performance against other attacks is much lower (e.g., 27.31\% for RA and 34.08\% for AdvDroidZero). KDE consistently lags behind, with accuracies $\leqslant$ 21.07\% on Drebin and $\leqslant$ 19.30\% on Androzoo.

Third, more advanced defenses like FD-VAE and PAD-SMA show better robustness, with FD-VAE reaching up to 73.74\% and PAD-SMA up to 70.37\% on Drebin dataset. However, their performance still falls short of MalPurifier, with gaps ranging from 17.76\% to 37.3\% across different attacks and datasets.

These results indicate that defenses such as adversarial training and auxiliary mechanisms, while effective against certain perturbation-based attacks, fail to generalize to more complex, structure-targeting threats. In contrast, MalPurifier achieves the highest accuracy across all scenarios. For black-box query attacks like MAB and EvadeDroid, the purifier effectively denoises the manipulative queries, yielding robust accuracies $\geqslant$ 90.88\% on both datasets. Even more impressively, MalPurifier demonstrates strong resilience against the white-box HRAT attack, achieving accuracies of 92.08\% on Drebin and 81.31\% on Androzoo, respectively.

In summary, the fact that MalPurifier achieves consistently high robustness under this setting when compared to existing methods, demonstrates that its purification framework is feature-agnostic and naturally generalizes to those attacks targeting structural and semantic features.

\begin{tcolorbox}[colback=gray!10,colframe=black,leftrule=0.5pt,rightrule=0.5pt,toprule=0.5pt,bottomrule=0.5pt,left=1pt,right=1pt,top=1pt,bottom=1pt]
\textbf{Answer to RQ7:} MalPurifier's effectiveness on graph-derived features confirms that its purification paradigm is not limited to binary feature spaces. It significantly outperforms other defenses, maintaining a robust accuracy of up to 92.38\% on Drebin and 94.38\% on Androzoo across all tested structural threats.
\end{tcolorbox}

\section{Discussion and Limitation}\label{section6}
While MalPurifier demonstrates strong performance, we acknowledge several limitations and avenues for future work. We additionally conduct experiments to assess the robustness of MalPurifier against the Mimicry attack, wherein the attacker introduces perturbations to a malware sample to closely resemble a benign application. Unfortunately, MalPurifier cannot effectively resist Mimicry ($\leqslant$ 60\%) and has less effectiveness as malware samples are guided by more benign samples. This is primarily because the DAE model tends to purify these samples as benign due to their similarity to benign samples, leading to their misclassification. To mitigate this issue, we believe that implementing countermeasures, such as generating a more diverse set of adversarial examples or enhancing the adversarial purification model itself, would be beneficial. We plan to explore these possibilities in our future research.

In the case of a white-box attack, where the adversary has comprehensive knowledge of both the model and the purifier, MalPurifier could potentially be circumvented. Since our approach focuses on purifying perturbations without modifying or re-training the detection model, an adaptive adversary might carefully design adversarial malware that can be free from purification. For example, the adversary might directly implant tiny malicious functions into benign software, making it challenging for the purification model to distinguish and remove them. Alternatively, an adversary could try to generate inputs that cause the DAE itself to malfunction or reconstruct the sample incorrectly in a way (e.g., output a very noisy or distorted sample) that confuses the downstream classifier. Future research could explore several avenues to bolster defenses against such adaptive threats, such as investigating multiple diverse purifiers or developing iterative purification schemes.

Another limitation of our approach may be the potential performance degradation as new malware samples and their adversarial variants evolve. For example, advanced attackers might systematically generate entirely novel perturbation algorithms unseen by the broader security community. Because our framework relies on mapping perturbed representations back to the manifold of known clean data, it intrinsically mitigates any custom perturbation as long as it deviates from the clean distribution. However, ensuring high resilience against perpetually evolving custom evasions remains an open challenge requiring continuous manifold updating. We believe that by incorporating new data and leveraging dynamic updates, we can continuously enhance the detection system's capabilities.

Although the proposed approach is designed for Android malware detection, the concept to eliminate perturbations and enhance the robustness of the detection system is not limited to the Android platform. We believe that our method can be  extended to a variety of different malware types (e.g., ELF or EXE binaries) by small modifications. For instance, the feature extraction process might need to be modified to suit the characteristics of the target platform. Furthermore, due to the differences in file structures, the level and distribution of adversarial perturbations will also be different. Appropriate parameter adjustments may be required for the generation of adversarial samples and the injection of protective noise in the purification mechanism in this paper. Future work could investigate the feasibility of extending this approach to different malware types and the necessary modifications to ensure its effectiveness.

While our approach may not be foolproof, we firmly believe that it substantially enhances the resistance of Android malware detection against diverse evasion attacks in a lightweight and plug-and-play manner. In addition, we believe that with further improvements and optimizations, our approach can be generalized to a wider range of adversarial attack scenarios.

\section{Related Work}\label{section7}
This section begins with a review of existing studies on ML-based Android malware detection methods, followed by an introduction to evasion attacks against these approaches and a brief discussion of state-of-the-art solutions.

\subsection{ML-based Android Malware Detection}\label{section6.1}
Researchers have developed numerous ML-based Android malware detection methods that typically classify APKs using features extracted from the manifest and bytecode. For instance, \emph{Drebin}\cite{arp2014drebin} identifies Android malware by exploiting binary static features and employing SVM for classification. \emph{MaMaDroid}\cite{onwuzurike2019mamadroid} extracts sequences of API calls and then trains classifiers like K-Nearest Neighbors (KNN) to detect malware.

Additionally, DL-based methods~\cite{kim2019multimodal,alzaylaee2020dl} have demonstrated remarkable capabilities. For example, Andre~\cite{zhang2019familial} is a hybrid representation learning approach that clusters Android malware from multiple sources and classifies them using a three-layer DNN when they behave like existing families. Qiu \emph{et al.}~\cite{qiu2022cyber} proposed a framework that extracts heterogeneous features and utilizes DNN to recognize unknown and evolving malware.

Given the widespread use and outstanding performance of DNNs in Android malware detection, this study aims to enhance the robustness of DNN-based detectors using an independently trained purifier to pre-process input samples. This purifier restores the feature representation of adversarial malware to its original version and preserves the features of clean samples as much as possible, enabling the DNN model to correctly classify Android applications.

\subsection{Evasion Attacks in Android Malware Detection}\label{section6.2}
In the context of Android malware detection, evasion attacks employ crafted inputs to mislead models such that malicious apps will be classified as benign. As discussed in Section~\ref{section2}, it can be divided into problem-space attacks and feature-space attacks.

Problem-space attacks modify the Android apps directly, such as perturbations onto Android manifest and Dalvik bytecode~\cite{chen2020android} or insertion of benign components into malicious samples~\cite{pierazzi2020intriguing}, for generating adversarial malware to deceive ML-based detectors. On the contrary, feature-space attacks map the malware example into a feature vector, and then introduce perturbations to the vector values~\cite{xu2023ofei} or reconstruct the vector representation~\cite{grosse2017adversarial} to achieve misclassification. Moreover, recent studies demonstrate that the utilization of ensemble attacks~\cite{dong2018boosting,li2020adversarial} intensifies the impact of the attacks, presenting a more formidable challenge for defense mechanisms.

To combat the escalating prevalence of evasion attacks, the method presented in this paper is not tailored to counter any specific attack. Instead, it strives to establish a universally applicable approach that effectively mitigates both problem-space attacks and feature-space attacks. Additionally, the proposed method significantly enhances robustness while maintaining accuracy on clean samples.

\subsection{Defenses against Evasion Attacks}\label{section6.3}
\emph{Adversarial training}~\cite{zhou2024mtdroid,lau2023interpolated,doan2023feature} is widely recognized as one of the most popular methods for defeating evasion attacks. Recent research~\cite{li2020adversarial,ficco2022malware} has further shown that combining adversarial training with \emph{ensemble learning} can enhance model robustness. However, it is worth noting that adversarial training typically retrains the model by generating and incorporating adversarial examples, which can lead to a significant increase in computational burden. Also, the defenses may not effectively mitigate attacks that differ significantly from the ones encountered during training.

In addition, there are also several countermeasures to identify evasion attacks through an auxiliary model. For example, Li \emph{et al.}~\cite{li2021robust} introduced a Variational AutoEncoder (VAE) to distinguish benign examples from adversarial malware according to reconstruction errors, and Li \emph{et al.}~\cite{li2023pad} leverages a convex DNN model-based detector to recognize the evasion attacks. Despite not requiring retraining of the target model, the auxiliary model remains closely coupled with the malware detection model and is still unable to effectively handle sophisticated and adaptive attacks.

Our work differs from prior methods in three key aspects. (i) Instead of training on a specific adversarial attack, MalPurifier employs diversified adversarial perturbations from no perturbation to worst-case scenarios to cover a broader perturbation landscape. (ii) To protect the recognizability of clean samples, MalPurifier actively injects protective noise during training to enhance the model's tolerance to variations between benign samples. (iii) Beyond detecting adversarial samples via threshold-sensitive reconstruction errors alone, MalPurifier adopts a dual-objective loss that jointly optimizes reconstruction fidelity and downstream classification alignment.

\section{Conclusion and Future Work}\label{section8}
In this paper, we addressed the critical problem of defending Android malware detectors against evasion attacks. Recognizing the unique challenges posed by the discrete feature space and diverse threat landscape, we introduced MalPurifier, a novel plug-and-play purification framework. Our key contributions include a diversified adversarial perturbation mechanism to enhance robustness against unseen attacks, and a protective noise injection strategy to maintain high accuracy on benign data. We designed a DAE-based purification model with a customized loss function that combines reconstruction and prediction objectives to optimize for both sample recovery and classification performance. Extensive experiments on two large-scale datasets against a set of evasion attacks, including perturbation-based and structure-based threats, demonstrate that MalPurifier significantly outperforms state-of-the-art defenses.

Given the current trend of rapid malware evolution in the wild, the future development of our work is to leverage dynamic update mechanisms into our current pipeline, enabling MalPurifier to adapt to emerging attack patterns without full retraining. Another future extension of our approach may be to investigate robust methods against poisoning attacks, including the purification samples in the both training and test phases. These two parts of the research will substantially improve the security of employing machine learning techniques in Android malware detection.

% or


\ifCLASSOPTIONcompsoc
  \section*{Acknowledgments}
\else
  \section*{Acknowledgment}
\fi

This work was supported in part by the National Natural Science Foundation of China under Grant No. 62202097 and No. 62072100, in part by the Frontier Technologies R \& D Program of Jiangsu under Grant No. BF2025026, and in part by the Jiangsu Funding Program for Excellent Postdoctoral Talent under Grant No. 2022ZB137.

\ifCLASSOPTIONcaptionsoff
  \newpage
\fi



%\bibliographystyle{IEEEtran}
\bibliographystyle{IEEEtran}
\bibliography{IEEEabrv,Reference}

%\begin{thebibliography}{1}
%\end{thebibliography}


\vskip -25pt
\begin{IEEEbiography}[{\includegraphics[width=1in,height=1.25in,clip,keepaspectratio]{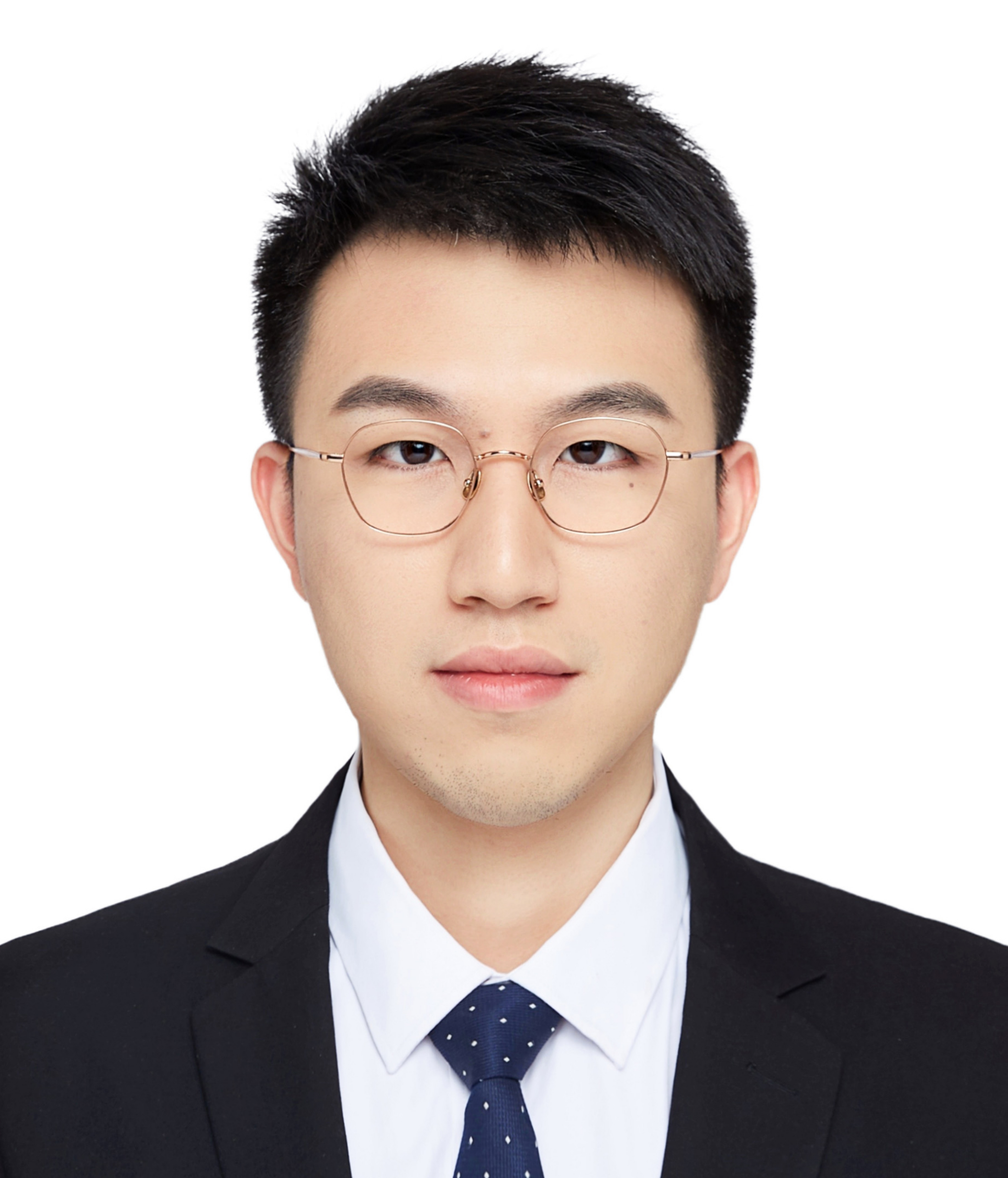}}]{Yuyang Zhou} is currently working as a postdoc with the School of Cyber Science and Engineering, Southeast University. His major research interests include Android malware detection, security modeling, and proactive DDoS mitigation. He has published in some of the topmost journals and conferences like IEEE TDSC, IEEE TIFS, IEEE TCCN, and ACM CCS, and is involved as a reviewer and in technical program committees of several journals and conferences in the field. He is a Member of IEEE and CCF.
\end{IEEEbiography}
\vskip-0.3in
\begin{IEEEbiography}[{\includegraphics[width=1in,height=1.25in,clip,keepaspectratio]{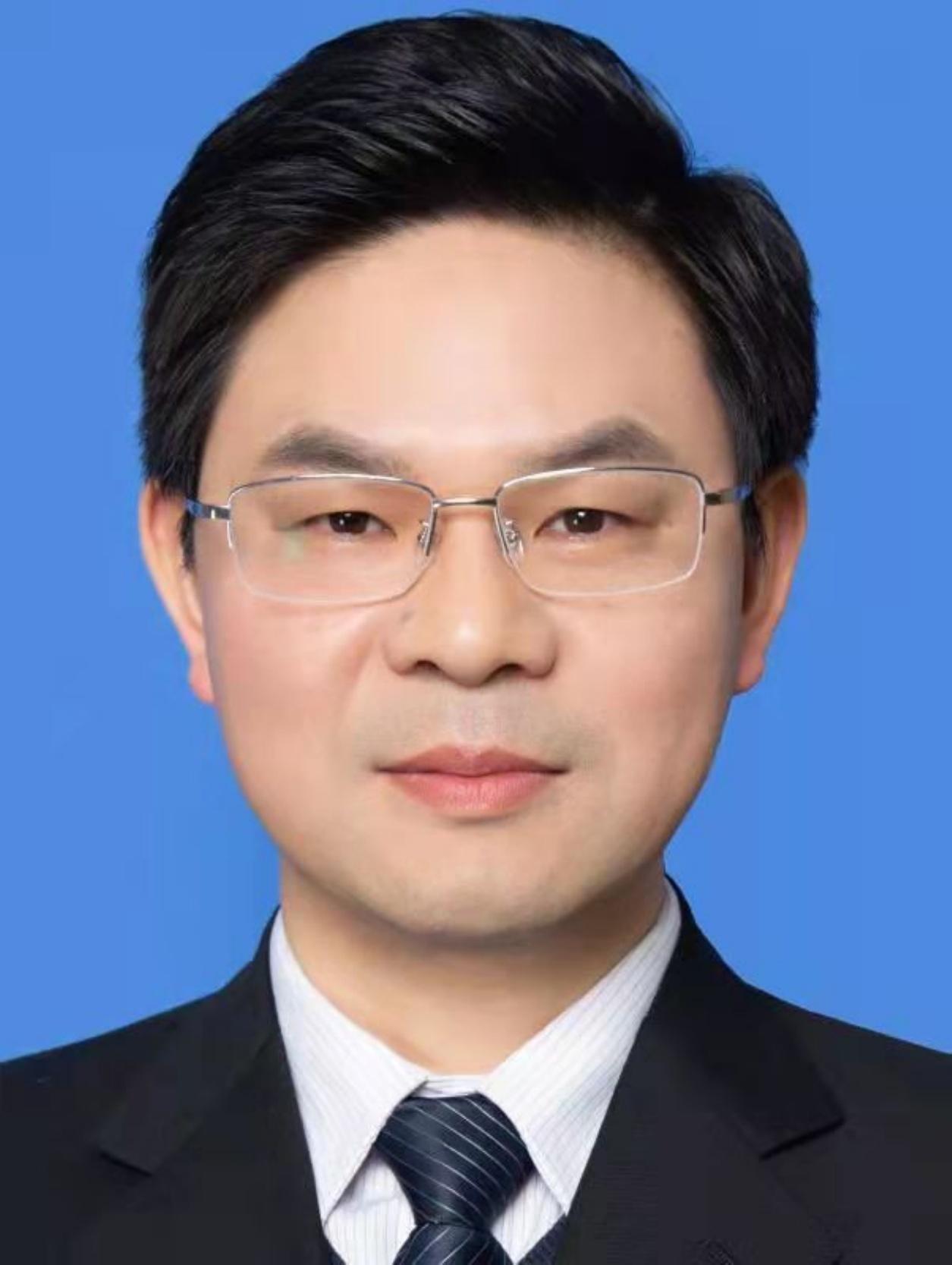}}]{Guang Cheng} is a Full Professor in the School of Cyber Science and Engineering, Southeast University, Nanjing, China. He has produced more than 100 technical papers, including top journals and top conferences like IEEE ToN, IEEE TIFS, IEEE TII, and INFOCOM. His research interests include network security, network measurement, and traffic behavior analysis. He is a Member of IEEE and a Distinguished Member of CCF. 
\end{IEEEbiography}
\vskip-0.3in
\begin{IEEEbiography}[{\includegraphics[width=1in,height=1.25in,clip,keepaspectratio]{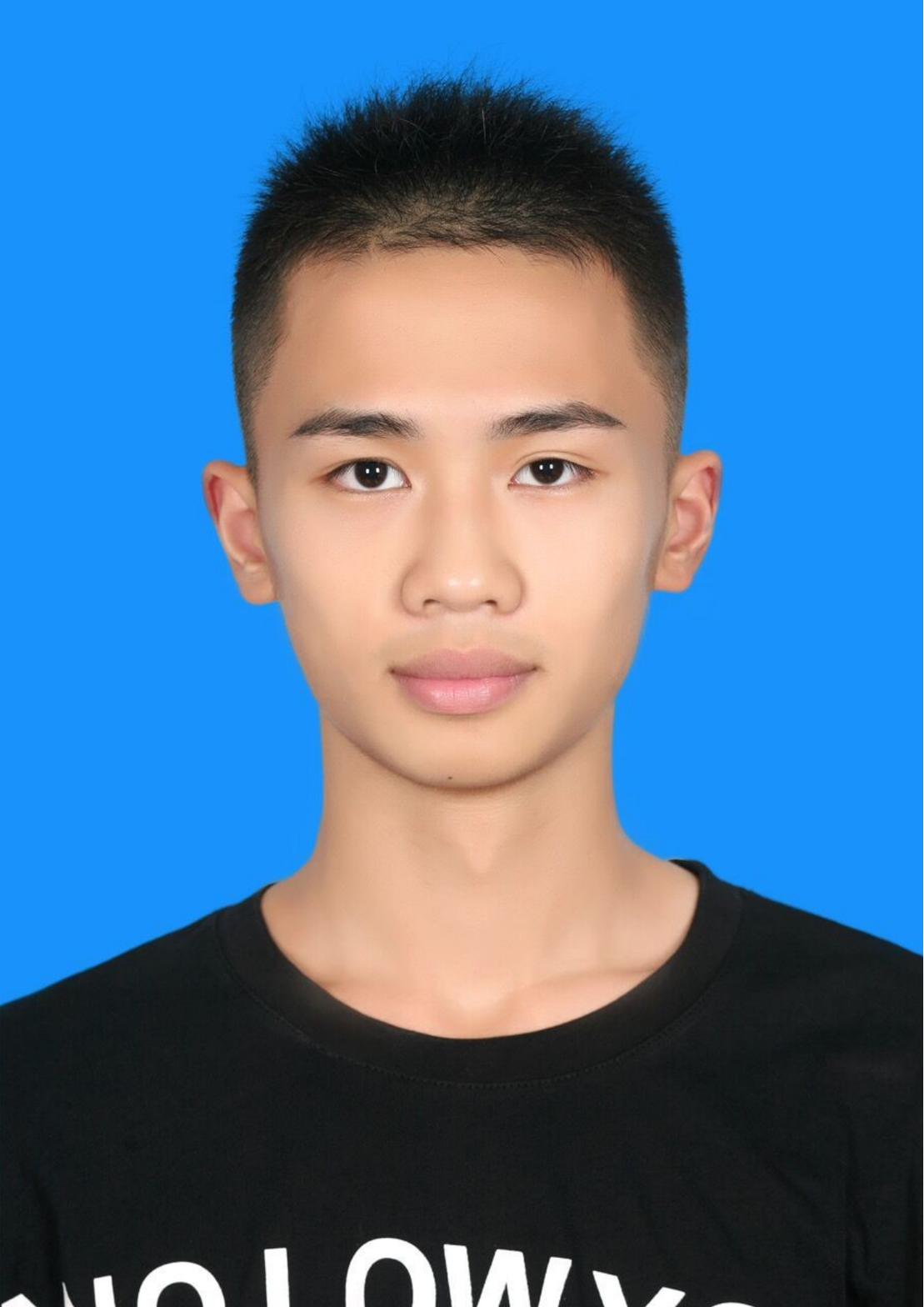}}]{Zongyao Chen} received the master degree in Cybersecurity from Southeast University in 2025. He is currently working at Alibaba (China) Co., Ltd. His major research interests include moving target defense, Android malware detection, and reverse engineering.
\end{IEEEbiography}
\vskip-0.3in
\begin{IEEEbiography}[{\includegraphics[width=1in,height=1.25in,clip,keepaspectratio]{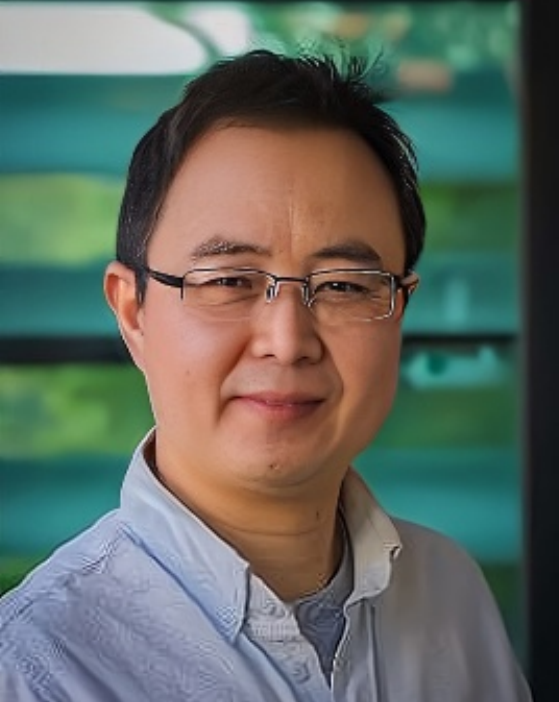}}]{Shui Yu} currently is a Professor of School of Computer Science, University of Technology Sydney, Australia. Dr Yu's research interest includes Big Data, Security and Privacy, Networking, and Mathematical Modelling. He has published two monographs and edited two books, and produced more than 500 technical papers, published in top journals such as IEEE TPDS, TC, TIFS, TMC, TKDE, TETC, ToN, and INFOCOM. He is a Fellow of IEEE.
\end{IEEEbiography}

%\newpage


%\vfill


\end{document}